\documentclass[prd,twocolumn,floatfix,preprintnumbers,reprint,showpacs]{revtex4}
\usepackage{graphicx}
\usepackage{dcolumn}
\usepackage{bm}
\usepackage{color}
\usepackage{amsmath}

\providecommand{\adsurl}[1]{\href{#1}{ADS}}

\def\lsim{\mathrel{\mathop
  {\hbox{\lower0.5ex\hbox{$\sim$}\kern-0.8em\lower-0.7ex\hbox{$<$}}}}}
\def\gsim{\mathrel{\mathop
  {\hbox{\lower0.5ex\hbox{$\sim$}\kern-0.8em\lower-0.7ex\hbox{$>$}}}}}

\def\CIV{\hbox{C$\,\rm \scriptstyle IV\ $}}
\def\SiIV{\hbox{Si$\,\rm \scriptstyle IV\ $}} 
\def\MgII{\hbox{Mg$\,\rm \scriptstyle II\ $}} 
\def\HeII{\hbox{He$\,\rm \scriptstyle II\ $}} 
\def\HI{\hbox{H$\,\rm \scriptstyle I\ $}} 
\def\mnras{Mon.\ Not.\ Roy.\ Astron.\ Soc.}

\begin{document}
\newcommand{\mincir}{\raise
-2.truept\hbox{\rlap{\hbox{$\sim$}}\raise5.truept 
\hbox{$<$}\ }}
\newcommand{\magcir}{\raise
-2.truept\hbox{\rlap{\hbox{$\sim$}}\raise5.truept
\hbox{$>$}\ }}
\newcommand{\minmag}{\raise-2.truept\hbox{\rlap{\hbox{$<$}}\raise
6.truept\hbox
{$>$}\ }}

\newcommand{\half}{{1\over2}}
\newcommand{\bk}{{\bf k}}
\newcommand{\Ocdm}{\Omega_{\rm cdm}}
\newcommand{\ocdm}{\omega_{\rm cdm}}
\newcommand{\OM}{\Omega_{\rm M}}
\newcommand{\OB}{\Omega_{\rm B}}
\newcommand{\oB}{\omega_{\rm B}}
\newcommand{\OX}{\Omega_{\rm X}}
\newcommand{\cltt}{C_l^{\rm TT}}
\newcommand{\clte}{C_l^{\rm TE}}
\newcommand{\mwdm}{m_{\rm WDM}}
\newcommand{\mnu}{\sum m_{\rm \nu}}
\newcommand{\etal}{{\it et al.~}}
 \newcommand{\lya}{{Lyman-$\alpha$~}}
\newcommand{\gad} {{\small {GADGET-2}}\,}
\input epsf

\title{Warm Dark Matter as a solution to the small scale crisis: new constraints from high redshift \lya forest data}

\author{Matteo
Viel$^{1,2}$,
George D.~Becker $^3$,
James S.~Bolton $^4$,
Martin G.~Haehnelt $^3$} 

\affiliation{
$^1$ INAF - Osservatorio Astronomico di Trieste, Via G.B. Tiepolo 11,
I-34131 Trieste, Italy \\
$^2$ INFN/National Institute for Nuclear Physics, Via Valerio 2,  
I-34127 Trieste, Italy \\
$^3$ Kavli Institute for Cosmology and Institute of Astronomy, Madingley Road, Cambridge, CB3 0HA, United Kingdom\\
$^4$ School of Physics and Astronomy, University of Nottingham, University Park, Nottingham, NG7 2RD, United Kingdom\\
}
\date{\today}

\begin{abstract}
We present updated constraints on the free-streaming of warm dark
matter (WDM) particles derived from an analysis of the \lya flux power
spectrum measured from high-resolution spectra of 25 $z > 4$ quasars
obtained with the Keck High Resolution Echelle Spectrometer (HIRES)
and the Magellan Inamori Kyocera Echelle (MIKE) spectrograph.  We
utilize a new suite of high-resolution hydrodynamical simulations that
explore WDM masses of 1, 2 and 4 keV (assuming the WDM consists of
thermal relics), along with different physically motivated thermal
histories. We carefully address different sources of systematic error
that may affect our final results and perform an analysis of the \lya
flux power with conservative error estimates.  By using a method that
samples the multi-dimensional astrophysical and cosmological parameter
space, we obtain a lower limit $\mwdm \gsim 3.3$ ~keV ($2\sigma$) for
warm dark matter particles in the form of early decoupled thermal
relics. Adding the Sloan Digital Sky Survey (SDSS) \lya flux power
spectrum does not improve this limit.  Thermal relics of masses 1 keV,
2 keV and 2.5 keV are disfavoured by the data at about the $9\sigma$,
$4\sigma$ and $3\sigma$ C.L., respectively. Our analysis disfavours
WDM models where there is a suppression in the linear matter power
spectrum at (non-linear) scales corresponding to $k=10\, h/$Mpc which
deviates more than 10\% from a $\Lambda$CDM model.  Given this limit,
the corresponding ``free-streaming mass'' below which the mass
function may be suppressed is $\sim 2\times10^8\,h^{-1}$ M$_{\odot}$.
There is thus very little room for a contribution of the
free-streaming of WDM to the solution of what has been termed the
small scale crisis of cold dark matter.

\end{abstract}


\maketitle
\section{Introduction}

The $\Lambda$CDM paradigm, in many respects, has proven to been an
immensely successful cosmological model.  $\Lambda$CDM is based on a
cosmological constant plus ``cold'' dark matter, i.e. dark matter
particles whose streaming velocities are negligible for most
astrophysical considerations. On large scales, the {\it Planck}
mission has just delivered another ringing endorsement of this model
with its first-year cosmology results \cite{planck}. On scales below a
few (comoving) Mpc, however, the matter power spectrum is still
difficult to probe, and it has been repeatedly suggested that dark
matter is perhaps ``warm'', with a free-streaming length that affects
the properties of low-mass (dwarf) galaxies.  Warm dark matter (WDM)
could alleviate the apparent difficulties of $\Lambda$CDM models in
reproducing some observations related to the matter power spectrum on
scales of a few Mpc and below. The most notable possible tensions
under $\Lambda$CDM are: the excess of the number of galactic
satellites, the cuspiness and high (phase space) density of galactic
cores, the luminosities of the Milky Way's satellites and the
properties of galaxies filling voids
(e.g.~\cite{bode,boylan,ferrero,weinberg}).

The main effect of the larger velocities of WDM particles, and the
resulting significant free-streaming length, would be to suppress
structures on Mpc scales and below.  The last few years have seen a
re-intensified discussion of this possibility, particularly in light
of improvements in numerical models and observations of the mass and
internal structure of Local Group satellites \cite{libeskind}. It has
been suggested that a free-streaming length corresponding to that of a
thermal relic WDM particle with a mass of 1-2 keV (and in some cases
as low as 0.5 keV) provides better agreement between the most recent
data and numerical simulations \cite{devega,lovell}. The difficulties
associated with the cold dark matter paradigm, however, arise on
scales where the matter spectrum is highly non-linear at $z \sim 0$,
and where very uncertain baryonic physics is known to play an
important role \cite{astrophy,wdm_nonlin,garrison}.

The \lya absorption produced by intergalactic neutral hydrogen in the
spectra of distant quasars (QSOs)--the so called ``\lya
forest''--provides a powerful alternative tool for constraining dark
matter properties, particularly the free-streaming of dark matter
particles on the scales in question. The \lya forest probes the matter
power spectrum in the mildly non-linear regime over a large range of
redshifts ($z=2-6$ in ground-based data) down to the small scales of
interest ($1-80\, h^{-1}$ Mpc) \cite{croftetal, Seljak:2006bg}.  
On large scales the SDSS-III BOSS collaboration has recently measured the
Baryonic Acoustic Oscillations (BAOs) scale in the 3D correlation
function of the \lya forest in $\sim$ 50,000 QSOs at $z\sim 2.2$
\cite{BAOs}. These findings further emphasize the value of the \lya forest as a
tracer of cosmological large-scale structure.  Constraints on the
matter power spectrum from \lya forest data on small scales are only
limited by the thermal cut-off in the flux power spectrum introduced
by pressure and thermal motions of baryons in the photo-ionized
intergalactic medium (IGM).  The IGM has a characteristic temperature
of $\sim 10^{4}\rm\,K$. While not trivial, modeling the relevant
physics with numerical simulations is reasonably straightforward; the
power spectrum at the relevant redshifts ($z\sim 2-5$) and scales is
only mildly non-linear and stellar feedback effects are much less
important than at lower redshifts \cite{vh06,feedback}.

The basic property of WDM, which impacts on both large scale structure
formation and the internal structure of dark matter haloes and the
galaxies they are hosting, is the significant ``thermal'' velocities
of the WDM particles (see \cite{bode}). The resulting
``free-streaming'' eliminates density fluctuations on scales below a
characteristic comoving wavenumber:
\begin{equation}
k_{\rm FS}\sim 15.6 \frac{h}{\mathrm {Mpc}} \left(\frac{\mwdm}{1 \mathrm{keV}}\right)^{4/3}\,\left(\frac{0.12}{\Omega_{\mathrm DM} h^2}\right)^{1/3},
\end{equation}
and leads to a very distinctive cut-off in the matter power spectrum
at a corresponding scale.  For example, the wavenumber at which the
linear WDM suppression reaches 50\% in terms of matter power, $k_{\rm
  1/2}$, w.r.t. the $\Lambda$CDM case can be approximated as:
\begin{equation}
k_{\rm 1/2}\sim 6.5 \frac{h}{\mathrm {Mpc}} \left(\frac{\mwdm}{1
  \mathrm{keV}}\right)^{1.11}\,\left(\frac{\Omega_{\mathrm
    DM}}{0.25}\right)^{-0.11}\,\left(\frac{h}{0.7}\right)^{1.22} \, ,
\end{equation}
where this equation uses the numerical results of
Ref.~\cite{Viel:2005qj}.  For standard thermal relics, the shape of
the cut-off is therefore well characterized in the linear regime and
there is an unambiguous relation between the mass of the thermal relic
WDM particle and a well-defined free-streaming length
(e.g.~\cite{Viel:2005qj}).  Note that we will also quote a
free-streaming mass, which is the mass at the mean density enclosed in
a half-wavelength mode corresponding to $k_{1/2}$.

For the sake of simplicity, the analysis here is presented in terms of
the mass of a thermal relic dark matter particle, for which there is a
one-to-one correspondence between the free-streaming length and the
particle mass.  We should point out, however, that in recent years
sterile neutrinos and other non-thermal particles have become popular
WDM candidates.  Some of these models are actually more similar to
mixed dark matter models with cold and warm dark matter components.
The shape of the free-streaming ``cut-off'' can then be quite
different from that of a thermal relic, and may instead correspond to
a downward step in the power spectrum rather than a cut-off (see
\cite{sterileboya,lovell}).  There is also no universal relation
between free-streaming length and mass of the WDM particles in these
models, and the normalization and functional form of this relation
varies greatly between different non-thermal WDM
candidates. Unfortunately, this has led to considerable confusion in
the literature when WDM models, characterized by their model-dependent
WDM particle masses, are compared between each other and/or thermal
relic models and in particular with \lya forest data.  For example
Ref.~\cite{lovell} quote a sterile neutrino mass of 2keV for their
thermal relic WDM model which corresponds, however, to a thermal relic
mass of 1.4 keV. For convenience and ease of comparison with the
literature, in this work we therefore consider only matter power
spectra with cut-off shapes expected for thermal relic WDM particles
and quote the unambiguous thermal relic masses (and corresponding
cut-off scale) to characterize our WDM models.

The Lyman-$\alpha$ forest, due to its spectral nature, probes the
matter power spectrum in velocity space. With increasing redshift the
ratio of a given (comoving) free-streaming length in velocity space to
the thermal cut-off length scale at a given temperature increases as
$\propto (1+z)^{1/2}$. There is furthermore strong observational
evidence that the temperature of the IGM decreases toward higher
redshift over the range $3<z<5$ \cite{becker2011a}, and therefore a
corresponding decrease in the thermal cut-off length scale. Despite
observational difficulties, pushing to high redshift allows the models
to probe smaller free-streaming lengths and thus to improve the limits
on WDM masses. The high-redshift regime has also the advantage of probing
structures that are more linear and the WDM cut-off is more prominent
at high redshift compared to low redshift \cite{wdm_nonlin}.

\lya forest data to constrain WDM were first used in Ref.~\cite{Nara}
where a limit of 750 eV was obtained by using N-body simulations only.
In previous work, Ref.~\cite{Viel:2005qj}, we used instead two
samples of high-resolution QSO \lya forest spectra at $z\sim2.5$ to
set a lower limit of 550 eV for the mass of a thermal WDM candidate.
Following this, Ref.~\cite{Seljak:2006qw} and Ref.~\cite{Viel:2006kd},
using higher-redshift QSO spectra from the Sloan Digital Sky Survey
(SDSS) and applying a different analysis method, significantly
improved this limit by a factor $\sim 4$.  As already noted, however,
care has to be taken in the correct modeling of the free-streaming
properties of ``non-thermal'' candidate WDM particles of a given
model-dependent mass, such as the popular sterile neutrino.  In
Ref.~\cite{sterileboya} the authors have focused on constraints on a
range of such models. Because of a non-zero mixing angle between
active and sterile flavour states, X-ray flux observations can also
constrain the abundance and decay rate of such WDM particles
(e.g.~\cite{xrayall}). The joint constraints from \lya forest data and
those from the X-ray fluxes of astrophysical objects now put
considerable tension on the parameter space allowed for a sterile
neutrino particle with the phase-space distribution proposed by
Dodelson \& Widrow \cite{dw,CDW}, although other, possibly more phsyical
scenarios should be explored \cite{sterileboya,kusenko}.

In Ref.~\cite{v08} we presented the most stringent \lya forest
limits up to that date on the free-streaming of dark matter, m$_{\rm WDM} >$ 4
kev (2$\sigma$).  That analysis was based on an (at the time) unrivaled
sample of high-quality, high resolution QSO absorption spectra
extending to $z\sim 5.5$. The limit is in obvious conflict with many
of the recent suggestions for alleviating the difficulties encountered
by numerical models in reproducing the observed properties of Local
Group satellite galaxies within the cold dark matter paradigm.  These
models often assume dark matter to be made up by thermal relic WDM
with masses in the range 0.5-2 keV (e.g.~\cite{lovell}).

Since our study in \cite{v08}, the size of our high-quality,
high-redshift QSO absorption spectra sample, the quality and size (in
particular the dynamic range and resolution) of our numerical
simulations  and our knowledge of the thermal and ionization
state of the IGM at the relevant redshifts have all significantly
improved. Motivated by these improvements, and in light of the lively
debate of dark matter possibly being warm with masses in the range
0.5-2 keV, we present here a new and much more extensive study of the
high-redshift \lya forest constraints on the free-streaming properties of dark
matter.  The new study is based on an improved data set, further
refined modeling of the flux power spectrum and a large suite of new
numerical hydrodynamical simulations.  We also perform a comprehensive
investigation of the the systematic uncertainties related to this
measurement.

Finally, it is worth highlighting that WDM would have profound implications
in many astrophysical and cosmological contexts.  In this respect, IGM
constraints are highly complementary to other probes based, for
example, on the properties of dark matter haloes \cite{maccio13}, the
number of satellites and their luminosities
\cite{boylan,macciofontanot}, strong lensing, the velocity function in
the local environment \cite{zavala}, phase-space density constraints
\cite{shi}, the formation of the first stars \cite{firststars}, the
high-redshift quasar luminosity function \cite{song}, decays of WDM
particles in the high redshift universe \cite{mapelli}, reionization
\cite{barkana}, gamma ray-bursts \cite{desouza}, galaxy formation
aspects \cite{menci} using N-body/hydrodynamical simulations
\cite{Wang:2007he,nonlinear} or analytical/semi-analytical methods
\cite{halomodel,valageas,pierpa,massfunc}.

The paper is organized as follows. In Section \ref{data} we present
our new data set. The simulations are described in Section
\ref{sims}. The mock quasar sample, which will be important for
estimating error amplitude and  covariance, is introduced in Section
\ref{mock}.  Section
\ref{fluxpower} discusses the effect of  the most important physical
parameters on the  flux power spectrum, while most of the remaining nuisance parameters and the impact they
have in terms of flux power are discussed in an Appendix.  Our main results are
reported in Section \ref{method}, together with a description of the
Monte Carlo sampling of the likelihood space.  We 
summarize our findings and  conclude in Section \ref{conclu}.

\section{Data}
\label{data}
Our analysis is based on high-resolution spectra of 25 quasars with
emission redshifts $4.48 \le z_{\rm em} \le 6.42$.  Compared to our
previous analysis in Ref.~\cite{v08} the number of QSO spectra, at
these redshifts, has improved by nearly a factor two. Spectra for
fourteen of the objects were taken with the Keck High Resolution
Echelle Spectrometer (HIRES) \cite{vogt1994}, and the remaining eleven
were taken with the Magellan Inamori Kyocera Echelle (MIKE)
spectrograph on the Magellan Clay telescope \cite{bernstein2002}.
Most of the data have been presented elsewhere
\cite{Becker,becker2011a,becker2011b,calverley2011}.  Here we briefly
review the relevant features of the spectra, and describe how the flux
power spectra were calculated.

The majority of spectra were reduced using a custom set of {\sc idl}
routines based on optimal sky subtraction \cite{kelson2003} and
optimal extraction \cite{horne1986} techniques, while a small subset
of the HIRES spectra (PSS~0248$+$1802 and BR ~1202$-$0725) were
reduced using the {\sc makee} software package.  The HIRES and MIKE spectra have spectral resolutions of 6.7 and
13.6~km\,s$^{-1}$ (FWHM), and the  
spectra were extracted using 2.1 and 5.0~km\,s$^{-1}$ spectral bins,
respectively.  The one-dimensional relative flux-calibrated spectra
were then continuum normalized using spline fits based on power-law
extrapolations of the continuum redward of the \lya emission line.  Median continuum
signal-to-noise ratios within the \lya forest of each object are
typically in the range of $10-20$ per pixel.  The continuum estimates are necessarily crude due to the high levels of absorption in the \lya forest at these redshifts.  We estimate that typical uncertainties in the continuum are of the oder $\sim$10-20\%, a point we return to in the power spectrum analysis.

To compute the flux power spectra, we first divided the \lya
forest in each quasar spectrum into two regions of equal redshift
length.  We then computed the power spectrum of the fractional
transmission, $\delta_F(z)$, in each region separately, where
\begin{equation}
\delta_{F}(z) = \frac{F(z) - \langle F(\bar{z}) \rangle}{\langle F(\bar{z}) \rangle}\, .
\end{equation}
Here, $\langle F(\bar{z}) \rangle$ is the mean transmitted flux
calculated at the mean redshift of each region.  We used fixed
relations for the mean flux given by $\langle F(z) \rangle =
\exp{[-\tau_{\rm eff}(z)]}$, where
\begin{equation}
\tau_{\rm eff}(z) =  
   \begin{cases}
   0.751\left( \frac{1+z}{4.5} \right)^{2.90} - 0.132, & z \le 4.5 \\
   2.26\left( \frac{1+z}{6.2} \right)^{4.91}, & z > 4.5 \, .
   \end{cases}
\end{equation}
The fit to $\tau_{\rm eff}$ at $z \le 4.5$ is from \cite{becker2013},
while the evolution at $z > 4.5$ is based on a fit to the mean flux
measured from the data presented here.  The latter is similar to the
trend in $\tau_{\rm eff}(z)$ presented by \cite{fan2006}.  We note that
our analysis is not sensitive to our choice of using a fixed relation
for $\langle F(\bar{z}) \rangle$.  In tests where we instead divided
each region by the mean flux in that region alone we obtained very
similar power spectrum estimates on average.  When calculating the
flux power spectrum we do not attempt to mask metal lines (see
discussion below).  We do, however, mask regions of strong telluric
absorption ($6275-6315$, $6865-6939$, $7594-7700$, $7163-7313$, and
$8126-8328$~\AA).  The longest-wavelength mask effectively means that
we probe up to a maximum redshift of 5.684.

The power spectra for individual regions were averaged over ten
logarithmic wavenumber bins in the range $\log_{10}k (\rm
s/km)=[-2.9,-1.1]$ with 0.2 dex spacing.  The power spectra from all
regions were then further averaged according to instrument and the
mean redshift in each region.  We used median redshifts
$z=4.2,4.6,5,5.4$ for a nominal total of eight combined power spectra
and 80 data points.  In order to be conservative, however, we decided
to use a subset of this sample and do not consider the highest
redshfit bin for the MIKE data set (which has very large error bars)
or the flux power measurements at $\log_{10} k (\rm s/km)<-2.3$, which
might be affected by continuum fitting uncertainties. The final data
set used in the present analysis thus consists of 49 data points.

Preliminary estimates of the error in the power spectra were
calculated using a bootstrap approach.  It is known, however, that
bootstrapping typical underestimates the true errors
(e.g. \cite{rollinde}).  To be conservative, we therefore decided to
add an additional 30\% uncertainty to our estimates of the errors of
the observed flux power spectrum for our standard analysis. We will
also quote the tighter limits that would be obtained without this
increase of the error estimate.  As a further check, we used the set
of mock QSO spectra described in Section \ref{mock} to determine what
the expected covariance in the power spectra should be (within the
limits of our finite simulation box) at each redshift for a sample of
similar size and quality to the one used here.  These estimate were
used to correct a few error estimates in the real data that appeared
to be too small.  With these corrections, the final flux power spectra
used here have error bars that are larger than $\sigma (P_{\rm
  F})/P_{\rm F} >0.075$.

\section{Cosmological hydrodynamical simulations} 
\label{sims}

We model the flux power spectrum based on a set of hydrodynamical
simulations performed with a modification of the publicly available
{\small{GADGET-II}} code.  This code implements a simplified star
formation criterion \cite{Springel:2005mi} that turns all gas
particles that have an overdensity above 1000 and a temperature below
$10^5$ K into star particles. This 
has been first used and extensively tested in Ref.~\cite{vhs04}.

The reference model, hereafter referred to as $(20,512)$, is a box of
length 20 $h^{-1}$ comoving Mpc with $2\times512^3$ gas and cold DM
particles (with a gravitational softening length of 1.3 $h^{-1}$ kpc)
in a flat $\Lambda$CDM universe with cosmological parameters
$\Omega_{\rm m}=0.274$, $\Omega_{\rm b}=0.0457$, $n_{\rm
  s}=0.968,H_0=70.2\rm\,km\,s^{-1}\,Mpc^{-1}$ and $\sigma_8=0.816$, in
agreement both with WMAP-9yr and Planck data \cite{wmap7,planck}.  We
further explore three different WDM models with masses $\mwdm=1,2,4$
keV; these models correspond to 50\% suppression of power in the
linear (redshift independent) matter power spectrum at scales $k_{1/2}
\sim 6.9,14.7,32\,h$/Mpc, respectively.  The initial condition power
spectra are generated with {\small CAMB}\cite{camb} and the
suppression and velocity for the WDM particles are implemented using
the approach outlined in Ref.~\cite{Viel:2005qj}.  In order to assess
convergence and evaluate resolution corrections (which are model
dependent), we also perform four additional $(20,768)$ models, one
each for the reference and WDM cases, and a single $(20,896)$ model
for the 2 keV simulation only. We also have performed three $(60,512)$
simulations for $\Lambda$CDM, WDM 1 keV and WDM 2 keV in order to
check the flux power spectrum convergence at the largest scales of our
smaller boxes.

We explore the impact of different thermal histories on the \lya
forest by modifying the Ultra Violet (UV) background photo-heating
rate in the simulations (e.g.~\cite{Bolton08}).  A power-law
temperature-density relation, $T=T_{0}(1+\delta)^{\gamma-1}$, arises
in the low density IGM ($1+\delta <10$) as a natural consequence of
the interplay between photo-heating and adiabatic cooling
\cite{HuiGnedin}.  We consider a range of values for the temperature
at mean density, $T_{0}$, and the power-law index of the
temperature-density relation, $\gamma$, based on the observational
measurements presented recently by Ref.~\cite{becker2011a}.  These consist
of a set of 3 different indices for the temperature-density relation,
$\gamma(z=4.6)\sim 1.0,1.3,1.6$, that are kept approximately constant
over the redshift range $z=[4.2-5.6]$ and 3 different temperatures at
mean density, $T_0(z=4.6)\sim 5400, 8300, 11200\,$K, which evolve with
redshift, yielding a total of 9 different thermal histories.  The
reference thermal history assumes
$(T_0(z=4.6),\gamma(z=4.6))=(8300\,\rm{K},1.3)$.  These 9 thermal
histories have been performed for all the 3 WDM models and for the
reference $\Lambda$CDM case, resulting in a total of 36 simulations.

In addition to these parameters we also consider and vary several
other physical parameters for the reference model only, given that
these are poorly constrained by the data.  These are the redshift of
reionization $z_{\rm re}$ (i.e. the redshift at which the optically
thin UV background is switched on in the simulations) which is chosen
to be $z_{\rm re}=12$ for the reference case and $z_{\rm re}=8,16$ for
two additional models; the Hubble constant, with two extra simulations
with $H_0=66.2,74.2\rm\,km\,s^{-1}\,Mpc^{-1}$; the scalar spectral
index, with $n_{\rm s}=0.968,0.998$; the matter content, with
$\Omega_{\rm m}=0.24,0.30$ and the r.m.s. amplitude of the matter
power spectrum, with $\sigma_8=0.77,0.87$.  We note here that varying
the redshift of reionization in particular enables us to assess the
impact of different integrated thermal histories (i.e. the effect of
Jeans smoothing, see \cite{rorai} for a recent discussion) on our
analysis.  The effect of different integrated thermal histories on
\lya forest constraints was also considered in Ref.~\cite{Viel09} using
this parameterisation.  Overall, a total of 54 hydrodynamical
simulations have been performed. Approximately 4000 core hours were
required for each (20,512) run to reach $z=2$, with the higher
resolution simulations requiring around 5 times longer.

During the simulation runs, we extract the non-linear matter power
spectra in order to compare with \cite{wdm_nonlin}.  For the reference
case only, we additioanlly extract the position of the haloes with a
friends-of-friends halo finding algorithm for our model of  the
impact of spatial fluctuations in the UV background on the flux power
(see the Appendix for further details).

\begin{figure}[h!]
\begin{center}
\includegraphics[angle=0,width=9.cm,height=9.cm]{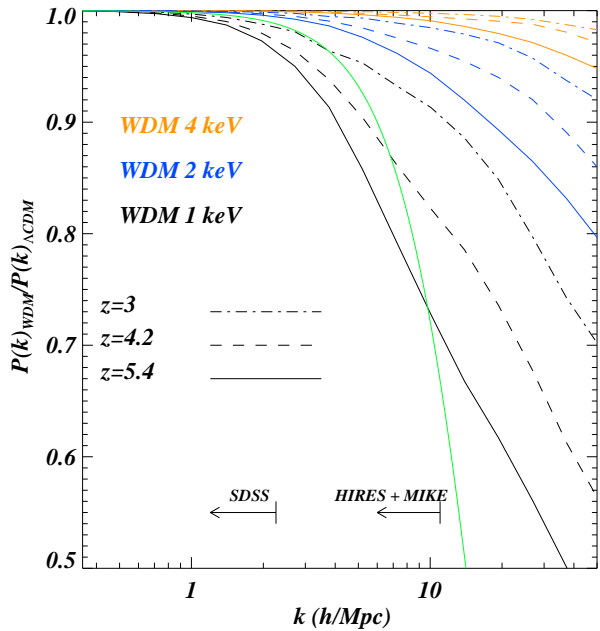}
\end{center}
\vspace{-0.5cm}
\caption{\label{fig_a} Ratio between the 3D non-linear matter power
  spectrum of 3 different WDM models (1, 2 and 4 keV, black, blue and
  orange curves) at 3 different redshifts ($z=3,\,4.2,\,5.4$,
  represented by the dot-dashed, dashed and continuous curves) and the
  corresponding $\Lambda$CDM model. The green curve represents the
  linear redshift independent suppression in terms of matter power for
  a $\mwdm=2$ keV model obtained using Eq.~6 of
  Ref.~\cite{Viel:2005qj}. The arrows in the bottom part of the figure
  indicate the maximum value of the wavenumbers probed by the SDSS
  data and by the data set used in the present analysis.  This figure
  refers to the reference (20,512) simulations.}
\end {figure} 

\begin{figure*}[!ht]
\label{fig_b}
\begin{center}
\includegraphics[angle=0,width=18.cm,height=5.cm]{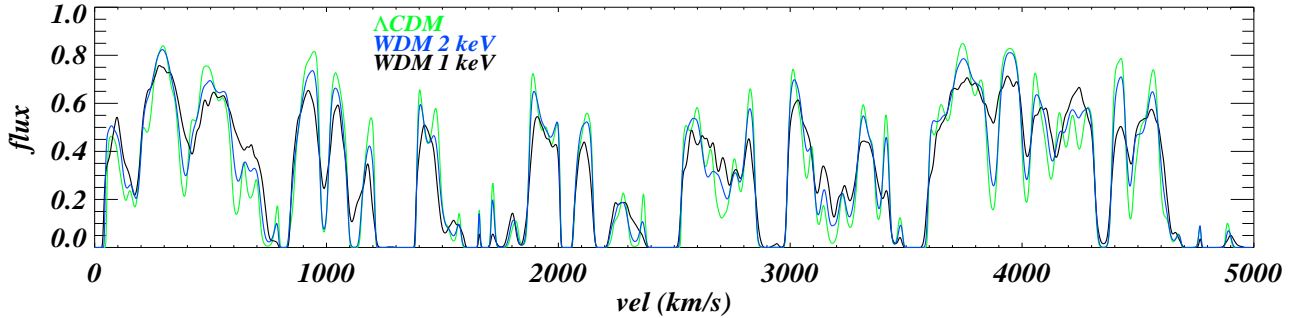}
\end{center}
\vspace{-0.5cm}
\caption{\label{fig_b} Transmitted flux along a set of random LOSs for
  the $\Lambda$CDM (green curve) and WDM 1 keV (black curve) and WDM 2
  keV (blue curve) models at $z=4.6$.  This figure refers to the
  reference (20,512) simulation without adding instrumental noise. The
  $\Lambda$CDM flux is clearly showing more substructure as compared to
  the WDM models.}
\end {figure*} 

Lastly, we note that the physical properties of the \lya forest
obtained from the {\sc TreePM/SPH} code {\sc GADGET-II} are in very
good agreement at the percent level with those inferred from the
moving-mesh code {\sc AREPO} \cite{Bird13} and with the Eulerian code
{\sc ENZO} \cite{Regan}.

\section{The mock QSO sample}
\label{mock}

The simulated \lya forest spectra are extracted along 5000 random
line-of-sights (LOSs) after interpolation of the relevant physical
quantities along the LOSs using the SPH formalism.  Box-size effects
on the flux power are estimated with $(60,512)$ simulations.  Note,
however, that the necesary box size correction is below the percent
level at the largest scales used. Resolution corrections are, however,
important.  The flux power spectra are corrected for resolution
effects using the $(20,768)$ simulations (see the Appendix for further
details).

The mean flux is varied {\it a posteriori}, after having extracted the
spectra, by reproducing $0.8,1,1.2$ times the observed $\tau_{\rm
  eff}$ (see Appendix). At the end of the procedure the four-dimensional parameter space in $(\mwdm, \tau_{\rm eff},T_0,\gamma)$ is explored
fully by means of quadrilinear interpolation performed over the set of
36 hydrodynamical simulations and 108 ($36 \times 3$ mean flux values)
flux models.

In order to get a better understanding of the expected (co)variance
properties of the observed data we generate samples of mock QSO
absorption spectra which resemble the observational data as closely as
possible.  The procedure used to create the mock spectra can be
summarized as follows: $i)$ we consider the total redshift path in
each redshift bin and combine the short simulated spectra ($20$
Mpc$/h$ in length) to match the total length of an observed QSO
spectrum (approximately 40 spectra are used); $ii)$ we allow for an
optical depth evolution along the LOS (which is absent since our
simulated spectra are from snapshots at fixed redshifts) following the
scaling expected from the fluctuating Gunn-Peterson approximation,
$\tau\propto (1+z)^{4.5}$ (see e.g. \cite{rauch97}); $iii)$ for each
short simulated spectrum we consider a $\pm$ 20\% error on the quasar
continuum placement (the continuum is drawn randomly from a Gaussian
distribution around the value 1 with a $\sigma=0.2$); $iv)$ we smooth
the flux with a Gaussian at a given FWHM corresponding to the
spectrograph resolution and rebin the spectra with the observed
pixel-size; $v)$ we add Gaussian-distributed noise on top of the flux,
matching the signal-to-noise of the observational data.  We
demonstrate in the Appendix that the (instrumental) effects of noise
and finite resolution, which are scale and redshift dependent, are
below 20\% (6\%) at the smallest scales for MIKE (HIRES). 

The (co)variance properties of this mock sample are in reasonable
agreement with those of our observed sample, both as a function of
redshift and wavenumber.  There are only 5 data points that appear to
have error bars that are smaller than those obtained from the mock
sample: 4 data points from the MIKE sample ($\log k$(s/km)$=-1.5$ at
$z=4.2$, $\log k$(s/km)$=-2.1$ at $z=4.6$, $\log k$(s/km)$=-1.3,-1.1$
at $z=5$) and one data point from the HIRES sample ($\log k=-1.1$ at
$z=4.2$). As the observed sample is still small and it is thus
expected that the bootstrap errors estimated from the data could be
unrealistically small, for these data points we increase the error
bars to match those obtained from a mock sample of 30 QSOs in the same
redshift bin.  These mock data points, with the extra 30\% error
added, otherwise agree well with those of the observational data,
giving us confidence that this is reasonable.

Finally, we note that the hydrodynamical simulations used to construct our
mock \lya forest spectra do not incorporate chemical elements other
than hydrogen and helium.  We have therefore estimated how
unidentified, lower redshift metal lines in the \lya forest may bias
our result, and in particular how these narrow absorption lines may
alter the flux power spectrum at small scales.  Furthermore, the
simulations also assume a spatially uniform UV background.  We
therefore also estimate the impact of spatial fluctuations in the UV
background on the \lya forest at $z=4.2$--$5.4$.  We follow a modified
version of the approach described in Ref.~\cite{BoltonViel11} for exploring
fluctuations in the \HeII ionising background at lower redshift, to
which we refer the reader for further details. These two systematics
effects will also be discussed more extensively in the Appendix.

\section{The flux power spectrum}
\label{fluxpower}
\subsection{The WDM and thermal cut-offs}

In this Section we demonstrate the distinctively different effects
that the thermal (i.e. due to the  temperature of the photo-ionized IGM) and WDM
cut-off have on the flux power spectrum.  We will also check for possible
effects due to the limited numerical resolution of our simulations.
Firstly, however, it is instructive to consider Fig.~\ref{fig_a},
where we show the ratio of the non-linear matter power spectra in the
WDM and $\Lambda$CDM simulations.  The results are shown at three
different redshifts, $z=(3,\,4.2,\,5.4)$.  The redshift range
$z=4.2$--$5.2$ brackets that of the high-resolution data set used in
this work.  We additionally present the matter power spectrum at $z=3$
to show the evolution of the non-linear power at lower redshift. The three
WDM models are reported as orange, blue and black curves for masses of
4,2 and 1 keV, respectively, while the green curve shows the
\emph{linear} suppression for the 2 keV case taken from
Ref.~\cite{Viel:2005qj}. The power spectra are already clearly
somewhat non-linear  at high redshift; the blue and green curves start
to differ significantly at small scales at $k>3
h/$Mpc. In the bottom part of the panel we show the approximate
wavenumber ranges that are probed by SDSS and the HIRES+MIKE data set
used in our analysis. Note that the non-linear matter suppression is
in good agreement with the fitting formula presented in
Ref.~\cite{wdm_nonlin}.  

In Figure \ref{fig_b} we qualitatively compare a set of noiseless
\lya forest spectra extracted from the $\Lambda$CDM, WDM 1 keV and WDM
2 keV models, represented by the green, black and blue curves
respectively.  It is clear that the amount of small-scale substructure
in the transmitted flux in the $\Lambda$CDM is more prominent with
respect to the WDM cases.  In the rest of this Section we will
quantify these differences in terms of the 1D flux power spectrum.

\begin{figure}[h!]
\begin{center}
\includegraphics[angle=0,width=9.cm,height=9.cm]{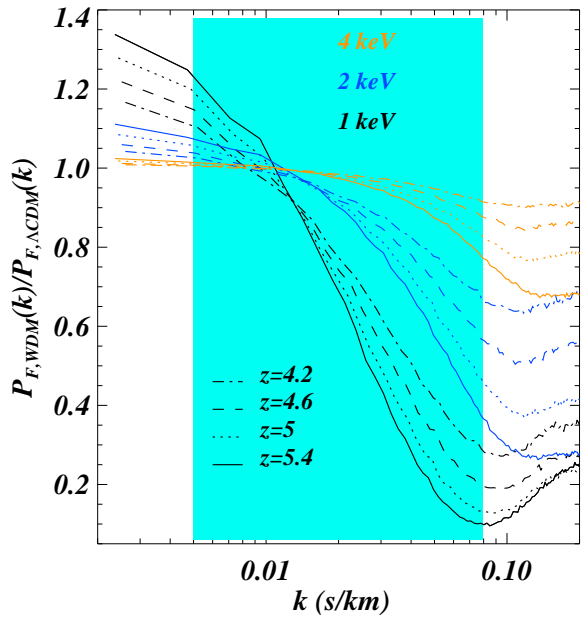}
\end{center}
\vspace{-0.5cm}
\caption{\label{fig_c} The ratio of the 1D flux power spectrum for 3
  different WDM models (1, 2 and 4 keV represented in black, blue and
  orange) at 4 different redshifts ($z=4.2,4.6,5,5.4$ represented by
  the continuous, dotted, dashed and dot-dashed curves, respectively)
  to the corresponding $\Lambda$CDM flux power spectra. This figure
  displays the results from the (20,512) simulations.  The mean flux
  is the same in all models and the shaded area shows the range of
  wavenumbers used in the present analysis.}
\end {figure}

We now turn to Figure \ref{fig_c}, which shows the ratio between the
1D flux power of the WDM and $\Lambda$CDM models for the four
different redshift bins used in the present analysis (note that we
compute the power spectrum of the quantity $\delta_{\rm F}=F/ \langle F \rangle - 1$, and we
refer to this as the flux power). The suppression of the flux power is
larger than that seen in the matter power spectrum. This is due to the
fact that the 1D matter power spectrum is an integral of the 3D power
spectrum and therefore very sensitive to the small scale cut-off.  As
expected, the largest differences exist between the 1 keV (black
curves) and the $\Lambda$CDM model. Note that the flux power also
changes at large scales; the requirement of reproducing the same
observed mean flux value (given by Eq. 4) results in an increase of
the power at those scales (the power spectrum of the WDM \emph{flux}
$F$, not $\delta_{\rm F}$, does show suppression over all scales when
compared to $\Lambda$CDM). Furthermore, we also note that there is a
substantial redshift evolution of the flux power between $z=5.4$ and
$z=4.2$.

\begin{figure}[h!]
\begin{center}
\includegraphics[angle=0,width=9.cm,height=9.cm]{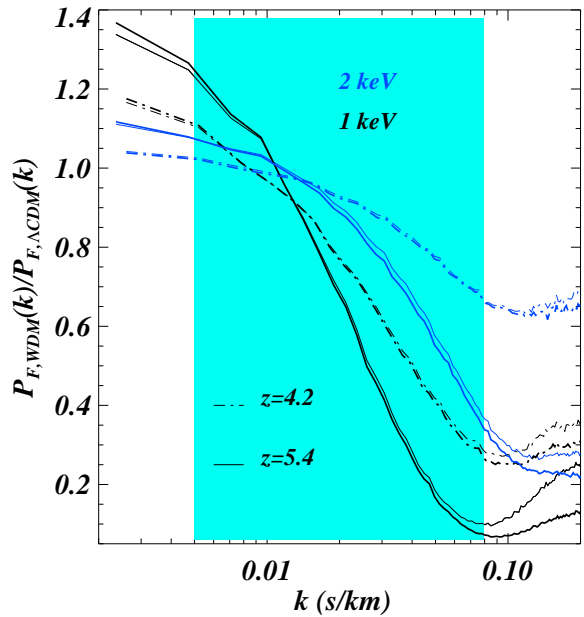}
\end{center}
\vspace{-0.5cm}
\caption{\label{fig_d} The ratio of the 1D flux power spectrum for 2
  different WDM models (1 and 2 keV, represented in black and blue) at
  2 different redshifts ($z=4.2,5.4$ represented by the continuous and
  dot-dashed curves, respectively) to the corresponding $\Lambda$CDM
  simulations. The thin curves refer to the (20,512) simulations,
  while the thick curves refer to the high resolution (20,768)
  models. The mean flux is the same for all models and the shaded area
  shows the range of wavenumbers used in the present analysis.}
\end {figure} 

Numerical convergence for WDM simulations can be particularly
difficult to achieve (see Ref.~\cite{Wang:2007he}). In Figure
\ref{fig_d} we demonstrate that at the resolution and WDM masses
considered in this work, this should, however,  not be an issue.  Figure
\ref{fig_d} compares the flux power extracted from the $(20,512)$ and
$(20,768)$ 1 and 2 keV simulations to the corresponding $\Lambda$CDM
simulations at the same resolution. The agreement between the
different resolution simulations is very good,   typically at the
percent level.  The differences are largest for the 1 keV case at the
smallest scales probed by our data in the current analysis, where they
reach the 10\% level.  The simulated flux power spectra for both
$\Lambda$CDM and WDM models have therefore been corrected for
resolution effects by multiplying the raw power spectra by the ratio of the results from the $(20,768)$ and $(20,512)$ simulations.  In general, we find the requirements for reaching
numerical convergence in terms of flux power are more demanding for
absolute values of the flux power rather than ratios of different
models w.r.t. the $\Lambda$CDM case. This will be discussed further in
the Appendix.

In Figures \ref{fig_f} and \ref{fig_g} we explore the effects of the
two thermal parameters, $T_0$ and $\gamma$, on the flux power spectrum.  As
discussed earlier, the $T-\rho$ relation is usually parameterized as a
power-law, $T(z)=T_0(1+\delta)^{\gamma-1}$.  In both of these figures
we also plot the WDM 2keV model in order to emphasize the very distinct
differences between the thermal and WDM cut-offs, both in the
dependence on wavenumber and redshift.  A
hotter (colder) $T_0$ value produces a suppression (enhancement) in
the flux power spectrum with a redshift dependent cut-off.  The WDM
cut-off is instead more pronounced and steeper than the cut-off
induced by a hotter IGM.  The dependence of the thermal cut-off on the
slope of the temperature-density $\gamma$ is Fig. \ref{fig_g} is also
very different from the wavenumber and redshift dependence of the WDM
cut-offs, and is much flatter over the wavenumber range considered
here.
\begin{figure}[h!]
\begin{center}
\includegraphics[angle=0,width=9.cm,height=9.cm]{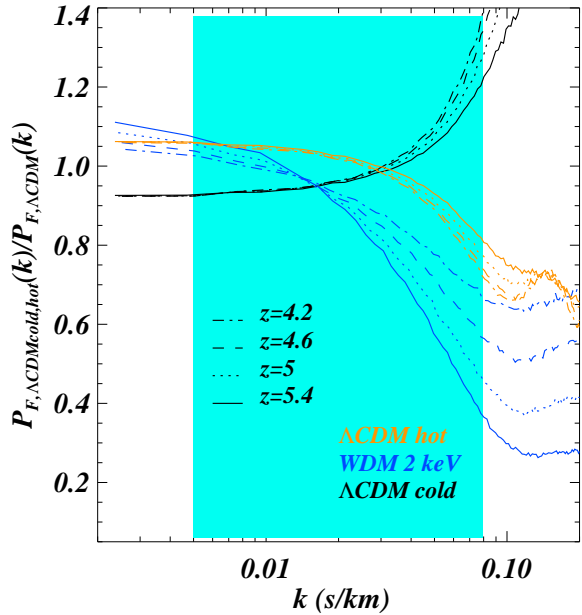}
\end{center}
\vspace{-0.5cm}
\caption{\label{fig_f} The ratio of the 1D flux power spectrum for two
  $\Lambda$CDM  models with different temperatures(HOT,  roughly hotter by 3000
  K with respect to the reference simulation, in orange and COLD, 
  roughly colder by 3000 K with respect to the reference simulation, in
  black) and at four different redshifts ($z=4.2,4.6,5,5.4$ represented
  by the dot-dashed, dashed, dotted and continuous curves,
  respectively) to the corresponding $\Lambda$CDM simulations. The WDM
  2 keV model is also shown in blue. The mean flux is the same for all
  models, and the shaded area shows the range of wavenumbers used in
  the present analysis.}
\end {figure}

\begin{figure}[h!]
\begin{center}
\includegraphics[angle=0,width=9.cm,height=9.cm]{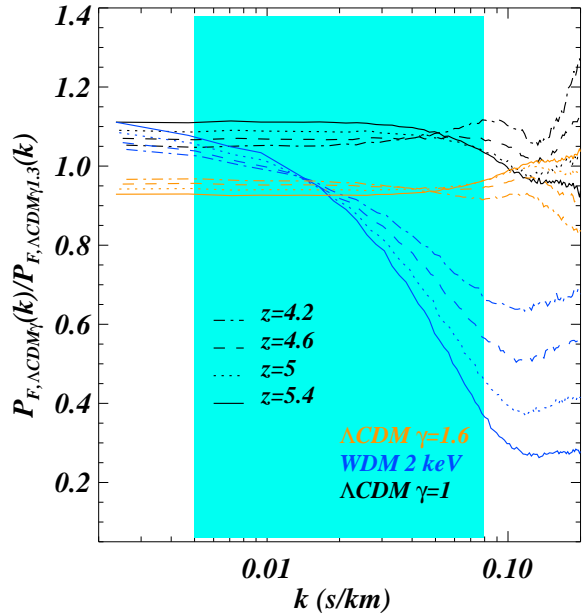}
\end{center}
\vspace{-0.5cm}
\caption{\label{fig_g} The ratio of the 1D flux power spectrum for two
   $\Lambda$CDM models with different slopes of the
  temperature-density relation ($\gamma=1.6$ in orange and
  $\gamma=1.0$ in black) and at four different redshifts
  ($z=4.2,4.6,5,5.4$, represented by the dot-dashed, dashed, dotted
  and continuous curves, respectively) to the corresponding
  $\Lambda$CDM simulations. The WDM 2 keV model is also shown in
  blue. The mean flux is the same for all models, and the shaded area
  shows the range of wavenumbers used in the present analysis. }
\end {figure}

The effects due to to changing the mean flux level are discussed in
detail in the Appendix (Fig.~\ref{fig_gg}). We conservatively assume
a range  of $\pm 20\%$ for the  observed effective optical depth. 
We further  note that the dependence of changing the mean flux level 
on wavenumber is even flatter 
than that obtained for variations of $\gamma$, but shows a weak scale
dependence 
in the highest redshift bin.

\subsection{Systematic uncertainties}

\begin{figure*}[!ht]
\begin{center}
\includegraphics[angle=0,width=15.cm,height=15.cm]{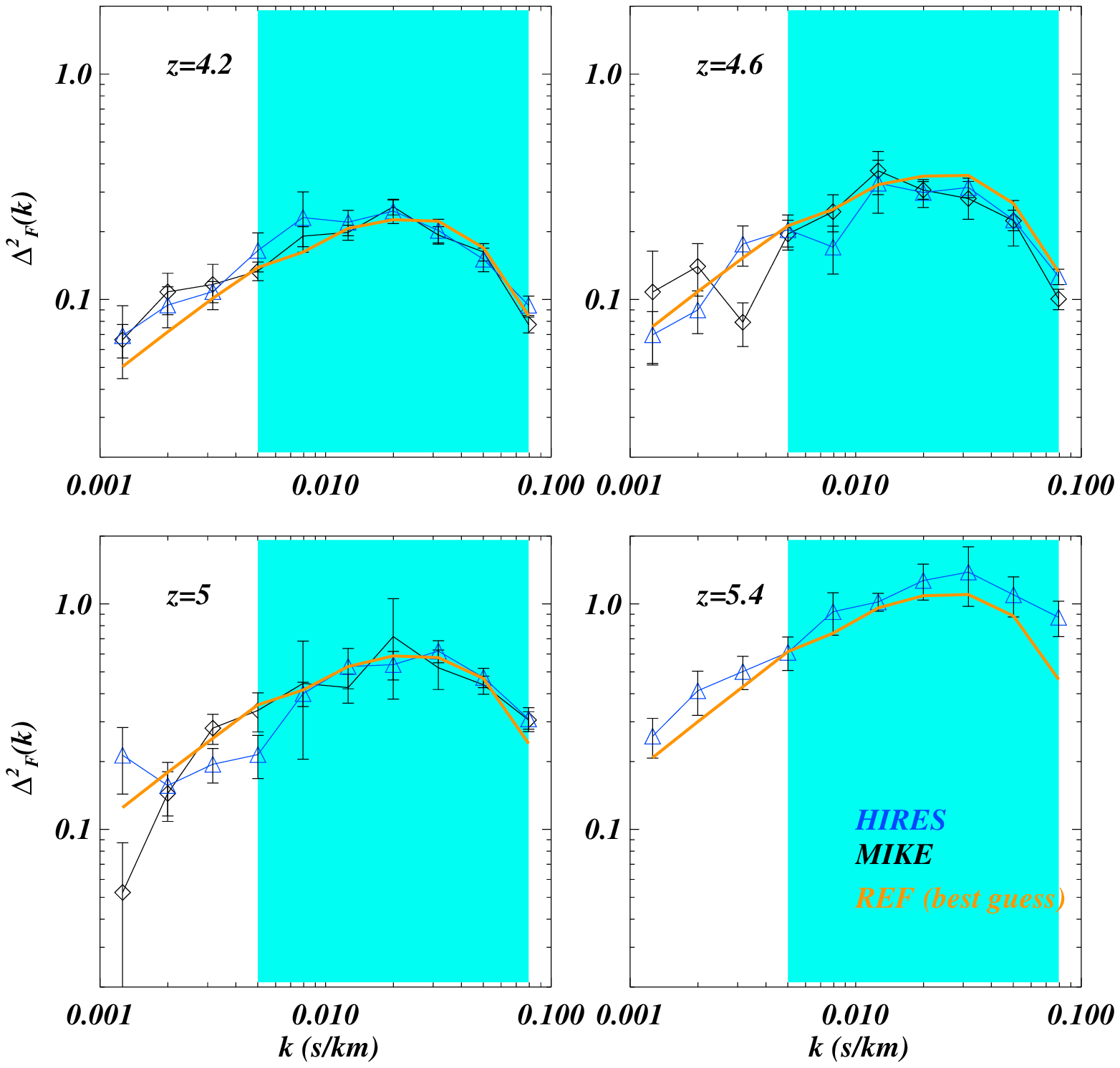}
\end{center}
\vspace{-0.5cm}
\caption{\label{fig_i} The flux power spectrum in dimensionless units,
  $P_F(k)\times k/!Mp!3$, used in the analysis
  performed. There are a total of 70 data points at 4 different
  redshifts. The reference $\Lambda$CDM model, which is our best guess
  starting point for the Monte Carlo Markov Chains, is also shown as
  orange curves. Only data points in the range $\log_{10}k (\rm
  s/km)=[-2.3,-1.1]$ are used in the analysis (shaded area).}
\end {figure*}

In this Section we now briefly discuss the following systematic
effects: instrumental resolution; noise; spatial fluctations in the UV
background and metal line contamination.  In the Appendix there is a
more detailed description of these nuisance effects and how they are
modeled.  We  only summarize the main quantitative results
here.  Instrumental resolution, which is different for the two
sub-data sets, suppresses the flux power spectrum by at most 20\% and 5\% for
MIKE and HIRES, respectively, at the smallest scales probed, with a
negligible redshift dependence. The signal-to-noise ratio impacts at
about the 2-3\% level at the smallest scales for $z\le 5$ while it is
at the 7\% level for the highest redshift bin.

The UV background fluctuations have been implemented with a
deliberately extreme model based on ionizing emission from quasars
only.  The impact of this extreme model of UV fluctuations on the flux
power spectrum is quite scale dependent, and rises considerably at
large scales (see also \cite{UV}). At the scales of interest here the effect on the flux
power spectrum is below the 10\% level (see Figure \ref{fig_h}).  This
should be considered as a generous upper limit.  The metal
contamination has a much smaller effect on the flux power spectrum,
below the 1\% level for the whole range of scales considered (see
Figure \ref{fig_hh}).  Apart from the apparently negligible metal
contamination, the other nuisance effects have been fully implemented
in our analysis.

In Table I, we summarize the main nuisance parameters and present
rough estimates of the relative errors induced in the flux power
spectrum.  Where possible, effects with a known amplitude such as
resolution and noise characteristics are simply incorporated into the
mock QSO spectra.  The remaining parameters are fully marginalized
over in our likelihood analysis.

\begin{table}[!ht]
\small
\caption{\small Summary of the estimates of the relative errors in  the flux
  power spectrum due to a range of  nuisance effects: the resolution of
  the observational data (the MIKE and HIRES data sets have different
  resolutions); the signal-to-noise ratio of the observational data;
  the numerical resolution of the simulations; contamination by metal
  absorbers at lower redshift; the mean flux level; the thermal
  history of the IGM and the fluctuations in the UV background. The
  table reports  estimates over the wavenumber range
  considered and the last three effects are properly marginalized over
  in the likelihood procedure.}
\label{tab0}
\begin{tabular}{lcc}
syst. eff. & $\sigma(P_{\mathrm F})/P_{\mathrm F}$ & Notes \\ 
\hline
\hline
\noalign{\smallskip}
data res. & $<5-15$\% & corrected \\
data S/N & $<3$\% & corrected\\
num. res. & $<5$\% & corrected \\
metals & $<1$\% & neglected \\
\hline
mean flux & $\sim 30$\% & marginalized\\
thermal history & $\sim 30$\% & marginalized\\
UV & $<10$\% & marginalized \\
\hline
\noalign{\smallskip}
\end{tabular}
\end{table}

\section{Method and Results}
\label{method}

We now turn to our analysis of the data and discuss our results.  For
all 108 flux models considered in our analysis, we compute the ratio
of these models with respect to the reference model.  We then bin this
ratio at the same wavenumbers as the data.  In practice, this means we
have a 4D parameter matrix with $(\mwdm,\,\langle F
\rangle,\,T_0,\,\gamma)$ that summarizes all the results obtained from
the hydrodynamical simulations, plus some further parameters for which
we have established the effect on the flux power for the reference
model only.  We parameterize the effect of UV background fluctuations
on the flux power with a factor $f_{\rm UV}$ that multiplies the flux
power spectrum corrections shown in Fig.~\ref{fig_hh}, constrained to
be in the range $[0,1]$ and applied in addition to the corrections
discussed in the previous Section ($f_{\rm UV}=1$ means that the power
spectrum is corrected exactly by the amount shown in the lower panel of
Fig.~\ref{fig_hh}). We decide to neglect the effect of metal
contamination since it is, as we discuss further in the Appendix, very
small.  We then perform second order Taylor interpolations for the
following remaining parameters: $z_{\rm reio},\Omega_{\rm
  m},\sigma_8,H_0,n_{\rm s}$, as in Refs.~\cite{Viel:2005ha,v08}.

In Figure \ref{fig_i} we compare our best-guess model (the reference
simulation), represented by the orange curves, with the observational
data. This best-guess model will be the reference point of our
likelihood code that will be described below. Note that there is
already rough ``visual'' agreement with the data, albeit with a poor $\chi^2$
value.

\begin{figure}[h!]
\begin{center}
\includegraphics[angle=0,width=6.cm,height=9.cm]{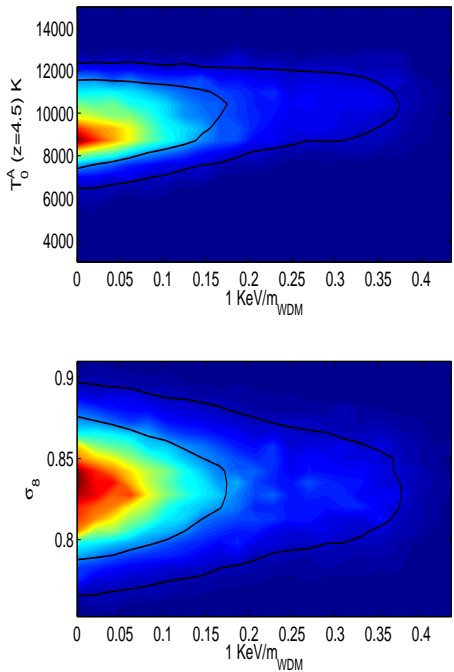}
\end{center}
\vspace{-0.5cm}
\caption{\label{fig2D_b} The two-dimensional 1 and 2$\sigma$ contours
  for mean (in colour) and marginalized (solid black curves)
  likelihoods for the parameters $1/\mwdm$ against $T_0^A(z=4.5)$ and
  $1/\mwdm$ against $\sigma_8$ obtained from the MIKE+HIRES data sets.
  These results assume a power-law evolution for $T_0$ and $\gamma$,
  but with $\gamma(z)$ constrained to be in the [0.7,1.7] range, and
  refer to a run for which some Planck-like priors on $\sigma_8$,
  $n_{\rm s}$ and $\Omega_{\rm m}$ have been applied.  Note, however,
  that our results are not sensitive to this choice of prior.}
\end {figure}

\begin{figure}[h!]
\begin{center}
\includegraphics[angle=0,width=9.cm,height=9.cm]{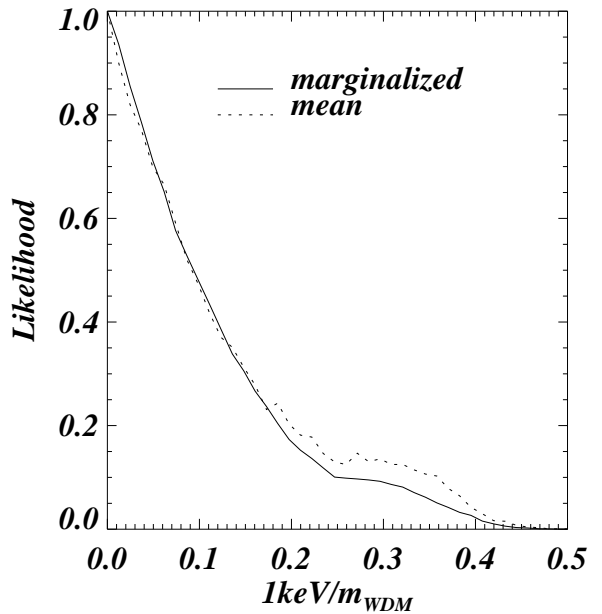}
\end{center}
\vspace{-0.5cm}
\caption{\label{fig1D} The one-dimensional mean (dotted curve) and
  marginalized (continuous curve) likelihoods for the parameter 1
  keV/$\mwdm$. These results refer to a run for which some Planck-like
  priors on $\sigma_8$, $n_{\rm s}$ and $\Omega_{\rm m}$ have been
  applied. Note, however, that our results are not sensitive to this.}
\end {figure}

We use a modified version of the code {\sc{COSMOMC}}
\cite{Lewis:2002ah} to derive parameter likelihoods from the \lya
forest data.  For the HIRES+MIKE data, we have a set of 15 parameters:
6 cosmological parameters ($\sigma_8, \Omega_{\rm m}, n_{\rm s}, H_0$,
$z_{\rm reio}$, $\mwdm$); 4 parameters describing the thermal state of
the IGM, using a power-law parameterization of the temperature-density
relation, $T=T_0(z)(1+\delta)^{\gamma(z)-1}$, with parameters
$T_0^A(z)=T_0^A[(1+z)/5.5)]^{T_0^S}$ and
$\gamma^A(z)=\gamma^A[(1+z)/5.5)]^{\gamma_A^S}$; 4 parameters
describing the evolution of the effective optical depth with redshift,
since a single power-law has been shown to be a poor approximation
over this wide redshift range (see \cite{Becker}) and one parameter
describing the spatial fluctuations in the UV background $f_{\rm
  UV}$. We apply strong Gaussian priors to $\sigma_8, \Omega_{\rm m},
n_{\rm s}$ in order to mimic Planck constraints: $\Omega_{\rm
  m}=0.315\pm0.017, \sigma_8=0.829\pm 0.013, n_{\rm s}=0.9603\pm
0.0073$. We have checked that these priors do not affect any of our
constraints on the free-streaming length/WDM mass, but they are
helpful in obtaining faster convergence of the
Monte-Carlo chains.  We vary $T_0^A$ in the range $[1000,20000]$ K and
$\gamma^A$ in the range $[0.7-1.7]$, and thereby heavily penalize the
$\chi^2$ if $\gamma(z=4.2,4.6,5,5.4)$ is outside the physical range
$[0.7-1.7]$. The values of $H_0$ and $z_{\rm reio}$ are not constrained
by the data and they are prior dependent: the range chosen are
$H_0=[50,100]$ km/s/Mpc and $z_{\rm reio}=[5,20]$, respectively.

The covariance matrix calculated from our data set is noisy
(especially at high redshift), preventing a reliable inversion. We
have therefore regularized the observed covariance matrix with the
correlation coefficients as estimated from the simulated spectra as in
Ref.  \cite{Lidz:2005gz}, $cov_{\rm d}(i,j)=r_{\rm
  s}(i,j)\sqrt{cov_{\rm d}(i,i)cov_{\rm d}(j,j)}$ with $r_{\rm
  s}(i,j)=cov_{\rm s}(i,j)/\sqrt{cov_{\rm s}(i,i)cov_{\rm s}(j,j)}$,
where $cov_{\rm d}$ and $cov_{\rm s}$ are the covariance matrices of
the observed and simulated spectra, respectively.

\begin{table}[!ht]
\small
\caption{\small Marginalized estimates (1 and
    2$\sigma$ C.L.) and best-fit values for a fit to MIKE+HIRES data
    using power-law fits for the evolution $\gamma(z)$ and $T_0(z)$. Planck
    priors on $\sigma_8$, $n_{\rm s}$ and $\Omega_{\rm m}$ have been
    applied. The best fit $\chi^2$ is 34 for 37 d.o.f. (49 data
    points - 12 free parameters) which has a probability of 39\% of
    being larger than this value.}
\label{tab2}
\begin{tabular}{lccc}
parameter & (1$\sigma$) & (2$\sigma$) & best fit\\ 
\hline
\noalign{\smallskip}
  $n_{\rm s}$      & [$0.942,0.97$]    & [$0.928,0.984$] & 0.957   \\
  $\sigma_8$    & [$0.806,0.856$]      & [$0.781,0.881$] & 0.822 \\ 
  $\Omega_{\rm m}$     & [$0.265,0.331$]    & [$0.234,0.362$] & 0.298 \\
  $\tau^A_{\rm eff} (z=4.2)$ & [$1.04,1.16$] & [$0.98,1.22$]& 1.16 \\ 
  $\tau^A_{\rm eff} (z=4.6)$  & [$1.19,1.33$]  & [$1.12,1.4$]& 1.32 \\ 
  $\tau^A_{\rm eff} (z=5)$& [$1.76,1.96$]  & [$1.66,2.05$] & 1.91 \\ 
  $\tau^A_{\rm eff} (z=5.4)$& [$2.72,3.06$] & [$2.55,3.21$]& 3.09 \\ 
  $\gamma^A (z=4.5)$       &  [$1.38,1.54$]&    [$1.09,1.65$]& 1.64 \\
  $\gamma^S (z=4.5)$    &  [$-0.76,1.1$]&        [$-2,2.3$]& -0.15 \\
  $T_0^A  (z=4.5) (10^3)$ K  &  [$9.1,10.4$]&    [$7.8,11.6$]& 9.2 \\ 
  $T_0^S  (z=4.5) (10^3)$ K  &  [$-3,-2.05$]  &  [$-3,-1.1$]& -2.5 \\ 
  $f_{\rm UV}$    & [$0-1$]  & [$0-1$] & 0.18 \\ 
  $z_{\rm reio}$  & [$5-11$]  & [$5-16.4$] & 11.2 \\
  $1$ keV/$m_{\rm WDM}$  & [$0-0.12$]  & [$0-0.3$] & 0.03 \\  
\hline
\noalign{\smallskip}
\end{tabular}
\end{table}

Our results are summarized in Table \ref{tab2}.  We obtain a $2\sigma$
upper limit on the parameter 1keV$/\mwdm$ of 0.3,which translates into
the following constraints: $\mwdm>3.3$ keV at the $2\sigma$ C.L. and
$\mwdm>8.33$ keV at the $1\sigma$ C.L., with a best-fit value of
$\mwdm=33$ keV.  For a $\sim 3$ keV WDM particle the 50\% suppression
in the 3D linear matter power compared to the $\Lambda$CDM case matter
power spectrum is at a scale of $k_{1/2}=22 h/$Mpc, while the
suppression at $k=10 h/$Mpc is about 10\%.  If we drop the 30\%
additional error applied to the observed flux power spectrum (see
Section II) we get a tighter lower limit of $\mwdm>4.5$ keV
($2\sigma$).  The $\chi^2$ gets worse by $\Delta \chi^2=14$, but still
has a probability of 11\% of being this large for the present number
of degrees of freedom.

We also took a ``frequentist'' approach and fixed the values of
$\mwdm$ to 2.5 and 3.3 keV and found the results in terms of the other
parameters: in this case the $\chi^2$ is of course higher than in the
$\Lambda$CDM model (with $\Delta \chi^2=5.6,3.8$, respectively) but
nevertheless compatible with the results obtained in our standard
analysis. This is similar to the approach used in Ref.~\cite{cwdm}, in
which an analysis of mixed cold and warm models was performed in both
a Bayesian and in a frequentist approach.

The degeneracies between the parameter 1keV$/\mwdm$ and the other
parameters are very weak. In Fig.~\ref{fig2D_b} we show the 2D contour
plots for the mean likelihood (in colour) and the marginalized
likelihood (black curves) for $T_{0}^{\rm A}$ and $\sigma_8$ versus 1 keV/$m_{WDM}$. In Fig.~\ref{fig1D} we report the 1D mean and marginalized 
likelihoods for 1keV$/\mwdm$ (continuous and dotted curves, respectively).

\begin{figure}[h!]
\begin{center}
\includegraphics[angle=0,width=9.cm,height=9.cm]{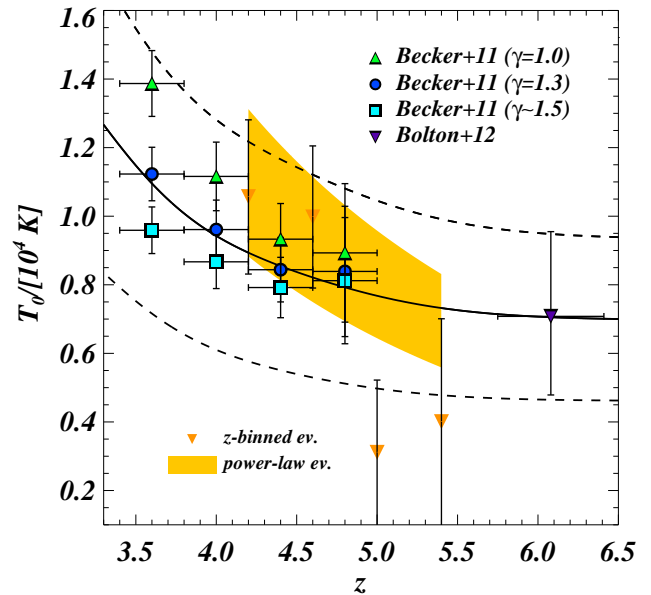}
\end{center}
\vspace{-0.5cm}
\caption{\label{fig_therm} The redshift evolution of the temperature
  at mean density, $T_0$, used in  our reference model is shown as
  continuous black curve, while the the two dashed line display our
  cold and hot models.  Recent measurements of the IGM temperature at
  mean density obtained by Ref. \cite{becker2011a} are also shown for
  different values of $\gamma$.  The measurement at $z\sim 6$ is
  taken from \cite{bolton12}.  The  results of our likelihood
  analysis are shown with the
  shaded orange area (for the power-law evolution case, $\pm 2\sigma$
  ranges), and with orange triangles for model where we left the
  temperature free in the four redshift bins
  ($1\sigma$ error bars).  In both cases the temperature values
  reported are the marginalized results.  These results refer to a run
  for which some Planck-like priors on $\sigma_8$, $n_{\rm s}$ and
  $\Omega_{\rm m}$ have been applied.}
\end {figure}

\begin{figure*}[!ht]
\begin{center}
\includegraphics[angle=0,width=18.cm,height=6.cm]{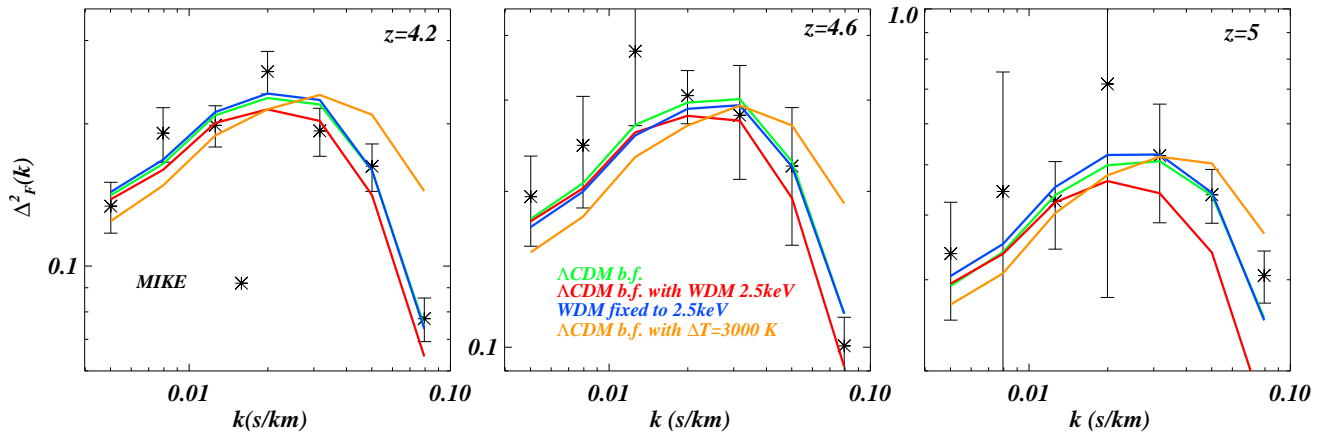}
\end{center}
\vspace{-0.5cm}
\caption{\label{figbf1} The best fit model for the MIKE data set
  (black crosses) used in the present analysis, shown as the green
  curves and labelled as ``$\Lambda$CDM b.f.''. This model is very
  close to $\Lambda$CDM.  We also show for qualitative purposes a few
  other models: a WDM model that has the same parameters as the best
  fit model except for the WDM mass (red curves) which is chosen to be
  2.5 keV; a model that has a hotter temperature (orange curves) 
  and a model for which the mass of the WDM is fixed to $\mwdm=2.5$
  keV, but for which all other parameters are set to their
  best-fitting values for this choice (blue curves).  Note that for
  the MIKE data we do not use the $z=5.4$ redshift bin.}
\end {figure*}

\begin{figure*}[!ht]
\begin{center}
\includegraphics[angle=0,width=12.cm,height=12.cm]{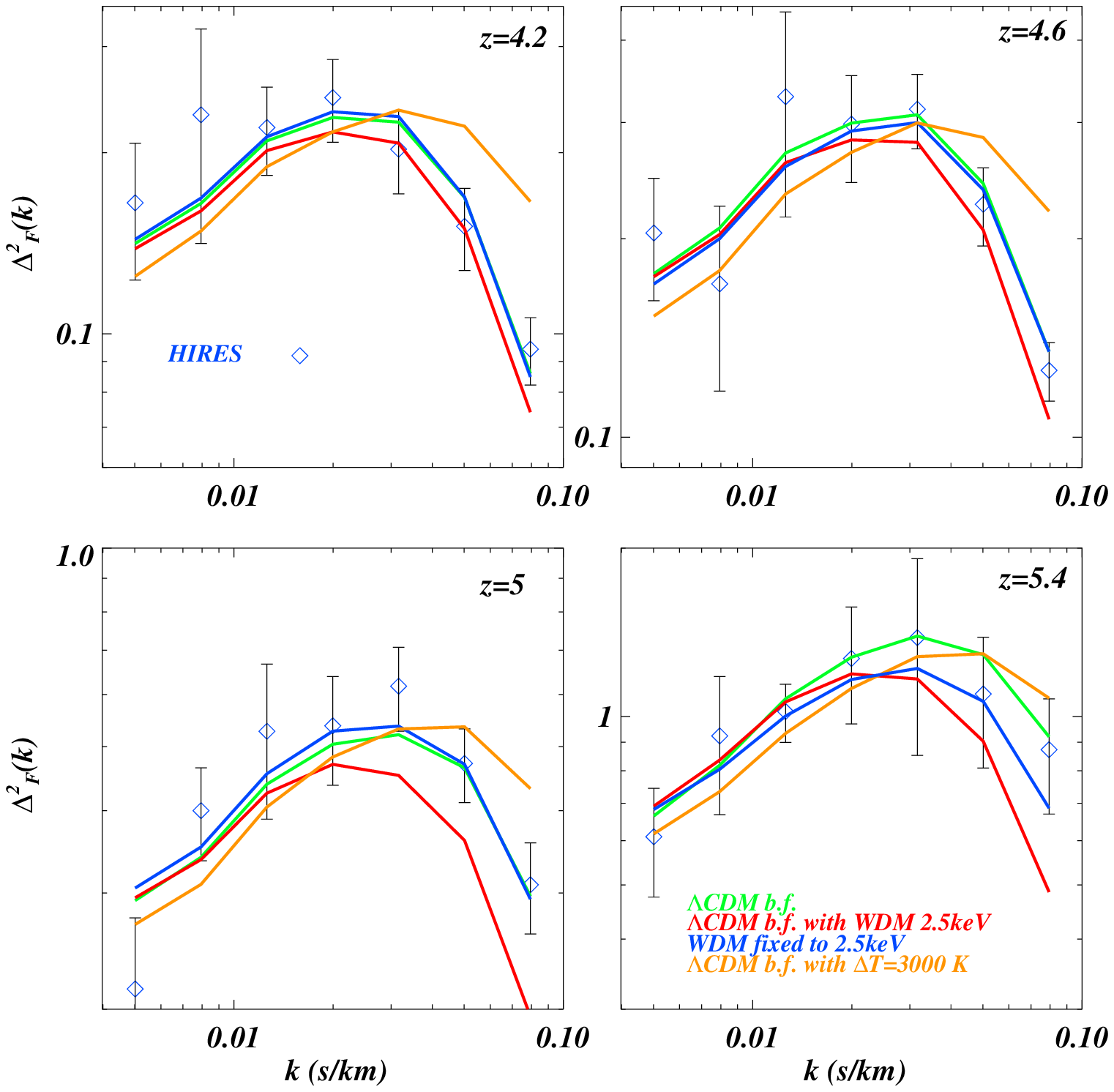}
\end{center}
\vspace{-0.5cm}
\caption{\label{figbf2} The best fit model for the HIRES data set
  (blue diamonds) used in the present analysis, shown as green curves
  and labelled as ``$\Lambda$CDM b.f.''.  As in Fig.~\ref{figbf1} we
  also show a few other models: a WDM model that has the same
  parameters as the best fit model except for the WDM mass (red
  curves) which is chosen to be 2.5 keV; a model that has a hotter
  temperature (orange curves) and a model for which the
  mass of the WDM is fixed to $\mwdm=2.5$ keV, but for which all other
  parameters are set to their best-fitting values for this choice
  (blue curves).  The last model, although visually similar to the
  $\Lambda$CDM model at $z \le 5$, is excluded at more than $2\sigma$
  confidence level.}
\end {figure*}

We  obtain the following evolution for the temperature-density
relation $T(z)=9200\,[(1+z)/5.5]^{-2.5}$ K and
$\gamma(z)=1.64\,[(1+z)/5.5]^{-0.15}$.  The inferred temperature is
decreasing with increasing redshift, while the redshift evolution of
$\gamma$ is weak.  We stress that the IGM thermal state is
  just one of several  nuisance parameter in our likelihood analysis
  over which we marginalise.  We
  discuss it here in the context of a consistency check rather than as
  a measurement.  With this in mind, in Figure \ref{fig_therm} we
show the recovered redshift evolution for $T_0$ compared to three
input thermal histories used in the simulations and measurements
obtained from high resolution \lya forest data from
Refs.~\cite{becker2011a,bolton12} (note that the power-law index of
the temperature-density relation, $\gamma$, has not yet been measured
directly at $z>4.2$).  The shaded orange
area brackets the $\pm 2\sigma$  of the temperatures 
obtained  by our standard likelihood analysis after
marginalization.  The inferred temperature evolution (parameterized  as a
power law in redshift)  is in good agreement with the measurements
from Ref.~\cite{becker2011a}.  
We also tested a model where the IGM temperature is left to vary freely in the
four redshift bins, shown by the orange data points with error bars in
Figure \ref{fig_therm}.  In the two highest redshift bins this 
analysis returns temperatures that are rather  cold and are disfavored by the data 
with an unreasonably  large temperature jump between $z=5$ and $z=4.5$
(see Fig.~\ref{fig_therm}). This suggests that in this case we have
introduced too many free parameters and are most likely
``overfitting'' the flux power spectrum.  For completeness we 
mention that with this apparently unphysical
temperature evolution our analysis gives a constraint on $\mwdm$
($2\sigma$ C.L.) which is about 1 keV lower than for our standard
analysis and also returns an unreasonably low reduced $\chi^2$ value.
Finally we have also performed a likelihood analysis where the IGM
temperature is fixed to be unrealistically  cold throughout (3000K, independent of
redshift) to allow for a maximum contribution of the free-straming of
WDM to the observed cut-off in the flux power. Again just for
completeness, for this model the constraint on $\mwdm$ ($2\sigma$
C.L.) is lower by about 0.5 keV compared to our standard analysis.

The recovered effective optical depth
values at each redshift bin are usually within 20\% of the measured
optical depth evolution used as the input into the likelihood
calculation. The inferred values for the amplitude and slope of the
matter power spectrum and for the matter content do not show biases
with respect to the Planck-like priors we used. Overall the $\chi^2$
for the best fit model is 34 for 37 d.o.f. which has a reasonably high
probability of about 60\% of being larger than this value.

Lastly, in Figures \ref{figbf1} and \ref{figbf2} we show our final
best-fit model compared to the data obtained with MIKE and HIRES,
respectively. The best fit model is shown as the green curves.  We
also overplot, for comparison purposes only, three other models that
are excluded with very high significance by the present analysis: a
model which has a WDM mass of $\mwdm = 2.5$ keV (red curves) and a hot
model with a temperature value which has been increased by 3000 K with
respect to the best fit case (orange curves).  
When calculating the
predicted flux power spectrum for these three models we change only
one parameter each time and leave all other parameters fixed at their
overall best fit values.  In addition, we show the case in which fix
$\mwdm = 2.5$ keV and allow all other parameters to assume their
best-fitting values under this assumption (blue curves).

Compared to our previous findings obtained in Ref.~\cite{v08}, it is
worth stressing the main differences.  First of all, from the data
side, the sample used here extends to high redshift and double the
amount of spectra contributing to the signal at $z>4$. Secondly, both
the simulations and the analysis have been refined by: increasing the
number of hydrodynamical simulations and their resolution; improving
the method in a way that allows a full sampling of the most relevant
parameter space (thermal parameters, WDM cutoff and mean flux)
compared to a poorer sampling of the parameter space made in
Ref.~\cite{v08}. When considering only the high-resolution data set, we
improve the limits by nearly a factor three from 1.2 keV to 3.3 keV at
the $2\sigma$ C.L., this is due to both the data and the modelling of
the flux power.

\section{Joint analysis with SDSS data}
In this Section we present the joint analysis with the Sloan Digital
Sky Survey (SDSS) 1D flux power spectrum data of
Ref.~\cite{McDonald:2004xn} where the authors have presented the flux
power spectrum of a sample of 3035 QSO absorption in the redshift
range $2 < z < 4$ drawn from the DR1 and DR2 data releases of
SDSS. These data have a spectral resolution of R $\sim 2000$, and so
typical \lya\ absorption features, which have a velocity width of
$\sim 30$ km\,s$^{-1}$, are not resolved.  The wide redshift range,
however, makes this data set very constraining in terms of
cosmological parameters.  As a final result of their analysis they
present an estimate of flux power spectrum $P_{\rm F}(k, z)$ at 12
wavenumbers in the range $0.00141 < k\,$ (s/km)$ < 0.01778$, equally
spaced in $\Delta\log k = 0.1$ for $z = 2.2, 2.4, 2.6, 2.8, 3, 3.2,
3.4, 3.6, 3.8, 4, 4.2$ for a total of 132 data points. This
measurement is likely to improve soon with the new analysis made by
the SDSS-III team of a sample which is about 50 times larger than the
one we utilize here \cite{palanque}.

\begin{figure*}[!ht]
\begin{center}
\includegraphics[angle=0,width=14.cm,height=14.cm]{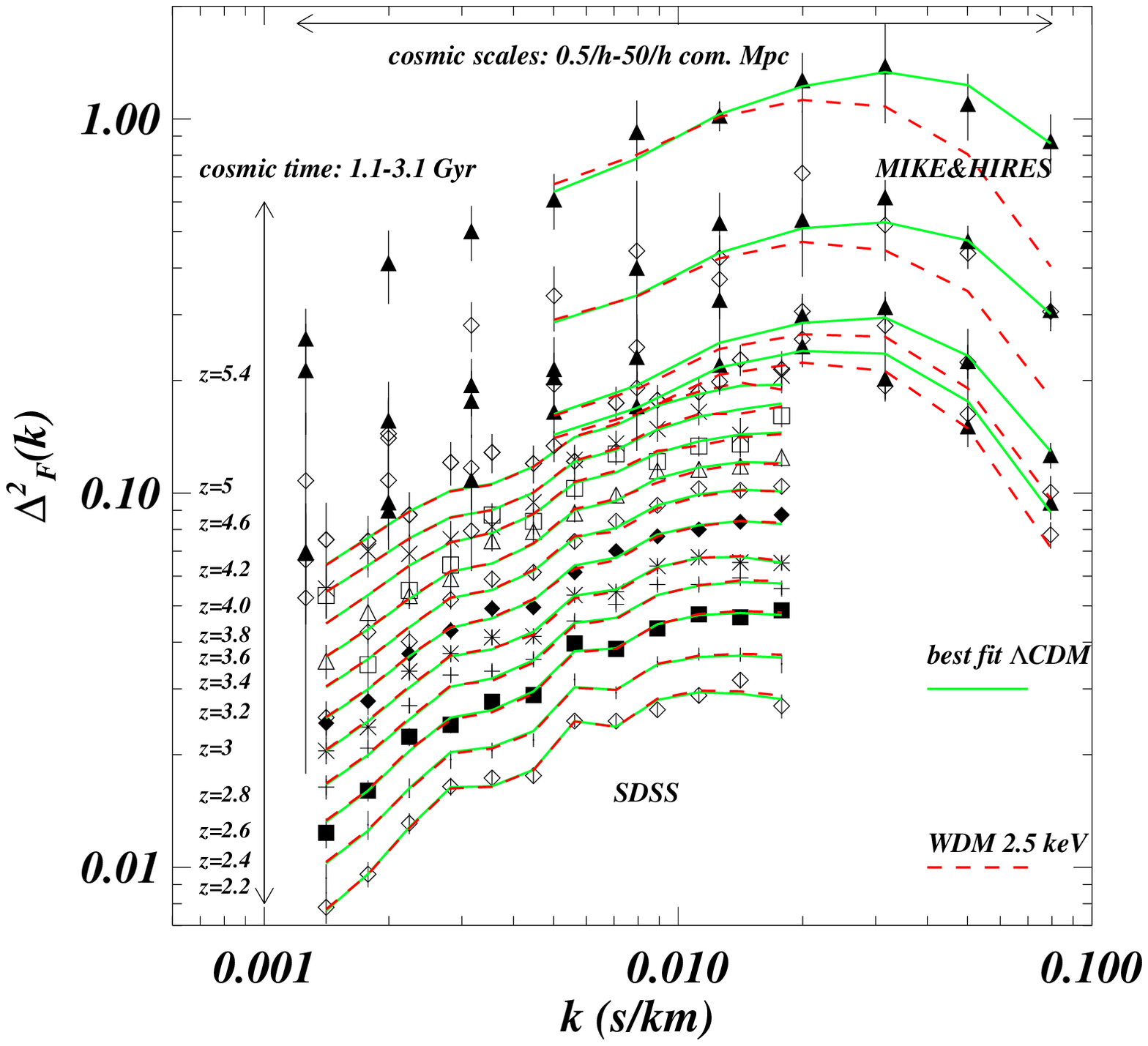}
\end{center}
\vspace{-0.5cm}
\caption{\label{all} Best fit model for the data sets used in the
  present analysis (SDSS+HIRES+MIKE) shown as green curves. We also
  show a WDM model that has the  best fit values of the green
  model except for the WDM mass (red dashed curves). These data span
  about two orders of magnitude in scale and the period 1.1-3.1 Gyrs
  after the Big Bang.}
\end {figure*} 

For the joint SDSS+MIKE+HIRES analysis we have used a total of 28
parameters: 15 parameters as used for the HIRES/MIKE spectra (without
$f_{ \rm UV}$ and the two parameters describing the effective optical
depth evolution at $z=5$) plus 13 noise-related parameters: 1
parameter which accounts for the contribution of Damped-\lya systems
and 12 parameters modeling the resolution and the noise properties of
the SDSS data set (see \cite{McDonald:2004xn}).  We do not consider
the possible effect of different reionisation scenarios on the SDSS
flux power. The covariance matrix of the SDSS flux power is provided
by the authors of \cite{McDonald:2004eu}. The $2\sigma$ lower limit on
$\mwdm$ is unchanged at 3.3 keV but now with a $\chi^2=183.3$ for the
best fit model for a total of 181 data points and 170 degrees of
freedom, which has a 23\% probability of being this large. Given the
fact that the SDSS and MIKE+HIRES data sets do not have redshift
overlap there is no significant bias in the other recovered
marginalized parameters.  Note that for the joint analysis we used the
SDSS likelihood based on second order Taylor expansion of the flux
power as described in Ref.~\cite{vh06} and used in Ref.~\cite{v08}.

Unlike our previous findings obtained in Ref.~\cite{v08}, where the
wide redshift range of SDSS data was helpful in breaking the
degeneracies between thermal parameters and WDM cutoff, we notice that
in this case the SDSS data do not improve the overall
constraints. This means that the constraining power of the new
high-resolution data set is higher than the low-resolution SDSS
data. The joint analysis gives now a lower limit of 3.3 keV, compared
to the previous 4 keV value is thereby slightly less stringent: this
is due to the different interplay between the data sets and to the
relative role of the degeneracies present between IGM thermal state
and WDM cut-off. It is also important to stress that the 30\% extra
error budget on the high-resolution data impacts on the final results
also for the joint analysis making the results less stringent than
in Ref.~\cite{v08}, where this error was not present.

\section{Discussion and Conclusions} 
\label{conclu}

\begin{table}[!ht]
\small
\caption{\small The final summary of the marginalized estimates (1 and
  2$\sigma$ C.L.) and best fit values for $\mwdm$. Planck priors on
  $\sigma_8$, $n_{\rm s}$ and $\Omega_{\rm m}$ have been applied. The
  REF. model refers to our reference conservative analysis; REF. w/o
  30\% refers to the case in which we do not add an extra 30\%
  uncertainty on the data to account for underestimated bootstrap
  error bars; REF. w/o covmat refers to the case in which we use only
  the diagonal terms of the covariance matrix; REF+SDSS is the joint
  analysis of our reference model and SDSS flux power.}

\label{tab3}
\begin{tabular}{ccccc}
model & (1$\sigma$) & (2$\sigma$) & best fit & $\chi^2$/d.o.f. \\ 
\hline
\noalign{\smallskip}
 REF.      & $> 8.3$ keV   & $ > 3.3$ keV  &  33 keV & 34/37  \\
 \hline
 \hline
 REF. w/o 30\%    & $>11.1$ keV  & $> 4.5$ keV &   100 keV  & 48/37 \\
 REF. w/o covmat  & $>7.7$ keV  & $>3.1$ keV &   14.3 keV &  33.2/37 \\ 
 REF. + SDSS   & $>7.2$ keV  & $>3.3$ keV &   42 keV &  183.3/170 \\
\hline
\noalign{\smallskip}
\end{tabular}
\end{table}

We have presented a comprehensive analysis of the transmitted \lya
flux power spectrum extracted from a set of 25 high-resolution QSO
spectra taken with the HIRES and MIKE spectrographs.  This represents
an improved and extended version of the sample originally analysed in
\cite{v08}.  The \lya forest is an excellent probe of the matter
distribution at intermediate and high redshift in the mildly
non-linear regime, from sub-Mpc up to BAO scales.  In this work we
have focused on constraining any possible suppression of the total matter power
spectrum which could be induced by the free-streaming of WDM particles
in the form of a thermal relic. Due to the non-linear nature of the
the relationship between the observed \lya flux and underlying
matter density, departures from the standard $\Lambda$CDM case are
expected over a range of scales that span at least one decade in
wavenumber space and can be constrained by the data used in the
present analysis.  We model this suppression by using a set of
high-resolution hydrodynamical simulations and by marginalizing over a
large range of physically motivated thermal histories. 

The WDM cut-off exhibits a distinctive behavior which we demonstrate
is not degenerate with other physical effects due to its different
redshift and scale dependence.  We consider possible sources of
systematic errors including metal line contamination, spatial
fluctuations in the UV background intensity and uncertainties in the
mean flux level estimation. Galactic feedback either in the form of
supernova driven galactic winds or Active Galactic Nuclei (AGN)
feedback should not impact the flux power spectra at the high redshift
considered in this analysis \cite{feedback}.

Our final results are obtained by means of a Monte Carlo Markov Chain
likelihood analysis around a best-guess reference model.  The constraints quoted
for $\mwdm$ have been calculated after marginalization over the other
astrophysical and cosmological parameters.  Our analysis is
conservative in the following sense: we have dropped the estimates of
the power spectrum at the largest scales probed by our sample in order
not to be sensitive to continuum fitting uncertainties; we add an
additional error of about 30\% to our error estimates obtained by
bootstrapping to account for the expected underestimation of the real
error; and we allow for large fluctuations in the UV background
fluctuations, which appear to be the most important nuisance factor.
Furthermore, we create a mock QSO sample which resembles as closely as
possible the real data including noise and resolution and use the
covariance matrix of this mock sample as an estimate of the error
properties of the real data.  Our final result of this analysis is
$\mwdm > 3.3$ keV at the $2\sigma$ C.L., where the mass refers to that
of a thermal relic. This mass implies that WDM models for which there
is a suppression in terms of the 3D linear matter power at scales
$k=10\,h$/Mpc ($k=22\,h$/Mpc) larger than 10\% (50\%) when compared to
the $\Lambda$CDM case, are disfavoured by the present data sets. The
corresponding value of the ``free-streaming'' mass is $\sim 2\times
10^8 M_{\odot}/h$.  A model with a 2.5 keV thermal relic mass is
disfavoured by the data at about $3\sigma$ C.L., a 2 keV mass at about
$4\sigma$ C.L., and a $\mwdm=1$ keV model at about $9\sigma$ C.L. Our
final marginalized estimates and best fit values for $\mwdm$ are
summarized in Table III.

Overall, the final results presented are similar to those we have
obtained in our previous analysis Ref.~\cite{v08} [3.3 (4.5) keV vs 4
  keV previously if we include (do not include) an additional 30\%
  error to account for a possible underestimate of the statistical
  error from a boot-strapping analysis). We emphasize, however, that
  the present analysis is considerably more robust. It uses a larger
  data set, a much improved analysis based on a broader suite of
  significantly improved simulations and as well as an extensive
  analysis of the systematic uncertainties.

Further improvement of the constraints on the free-streaming of dark
matter particles from \lya forest data could come mainly from an
enlarged set of high-quality, high resolution spectra, especially at
the highest redshifts where the flux power spectrum is most sensitive
to the free-streaming of dark matter.  An increase of the dynamical
range of the simulations and improved independent constraints on the
thermal state and thermal history of the IGM are next on the list as
requirements for further corroborating and perhaps pushing the
constraints to even larger thermal relic masses.  In the future,
considerably stronger constraints on WDM may be derived using a
baryonic tracer which is colder than the photo-ionized IGM, thus
moving the thermal cutoff to smaller scales in the flux power
spectrum.  Studies of 21 cm absorption/emission by neutral hydrogen
gas before reionization, for example, could eventually fulfill this
requirement.

However, with a lower limit of $10^{8} $M$_{\odot}\,h^{-1}$ for the
mass of dark matter haloes whose abundance could still be
significantly affected, the \lya forest data appears to leave already
very little room for a contribution of the free-streaming of warm dark
matter to the solution of what has been termed the small scale crisis
of cold dark matter. In particular, recent suggestions for models with
relic masses of 0.5-2keV are significantly disfavoured by our
analysis.  We finally note that our analysis also suggests that it is
unlikely that  sterile neutrinos could act in that role.

\section*{Acknowledgements}   
A part of the observations were made at the W.M. Keck Observatory
which is operated as a scientific partnership between the California
Institute of Technology and the University of California; it was made
possible by the generous support of the W.M. Keck Foundation.  This
paper also includes data gathered with the 6.5 meter Magellan
Telescopes located at Las Campanas Observatory, Chile.  The authors
wish to recognize and acknowledge the very significant cultural role
and reverence that the summit of Mauna Kea has always had within the
indigenous Hawaiian community.  We are most fortunate to have the
opportunity to conduct observations from this mountain.  The
hydrodynamical simulations in this work were performed using the
COSMOS Supercomputer in Cambridge (UK), which is sponsored by SGI,
Intel, HEFCE and the Darwin Supercomputer of the University of
Cambridge High Performance Computing Service
(http://www.hpc.cam.ac.uk/), provided by Dell Inc. using Strategic
Research Infrastructure Funding from the Higher Education Funding
Council for England. COSMOS and DARWIN are part of the DIRAC high
performance computing facility funded by STFC.  MV is supported by the
FP7 ERC grant ``cosmoIGM'' GA-257670, PRIN-MIUR and INFN/PD51 grants.
GDB acknowledges support from the Kavli foundation.  JSB acknowledges
the support of a Royal Society University Research Fellowship.  MH
acknowledges support by the FP7 ERC Grant Emergence-320596.

\appendix*
\section{Systematics}
\subsection{Numerical Convergence}

\begin{figure}[h!]
\begin{center}
\includegraphics[angle=0,width=9.cm,height=9.cm]{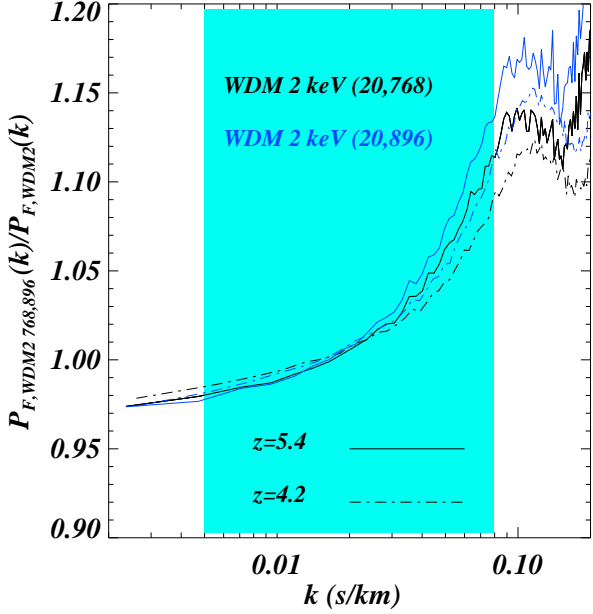}
\end{center}
\vspace{-0.5cm}
\caption{\label{fig_e} The ratio of the 1D flux power spectrum for two
  WDM 2 keV simulations at different resolutions, (20,768) and
  (20,896) represented by the black and blue curves respectively, to
  the reference WDM 2 keV run (20,512) at two redshifts ($z=4.2,5.4$
  represented by the continuous and dot-dashed curves,
  respectively). The mean flux is the same in all the models shown,
  and the shaded area indicates the range of wavenumbers used in the
  present analysis.}
\end {figure}

In Figure \ref{fig_e} we compare the 1D flux power spectrum extracted
from the WDM 2 keV $(20,768)$ and $(20,896)$ models to our reference
resolution of $(20,512)$.  We focus on this WDM mass since it is
excluded at high significance by our analysis, yet it is used by the
large scale structure community in order to solve apparent problems of
$\Lambda$CDM at small scales.  There is at most a 10-12\% correction
at the smallest scales probed in our analysis ($k\sim 0.08$ s/km) when
the flux power spectra from the $(20,768)$ model are considered, and
there is an extra 2\% correction when the $(20,896)$ is considered.
After correcting for this, we therefore believe that we have reached
5\% agreement in terms of the flux power for the smallest scales
considered in this work.  This is below the 7.75\% (1$\sigma$)
statistical error of the data at the same wavenumber.

\subsection{Instrumental Effects on the Flux Power}

\begin{figure}[h!]
\begin{center}
\includegraphics[angle=0,width=9.cm,height=9.cm]{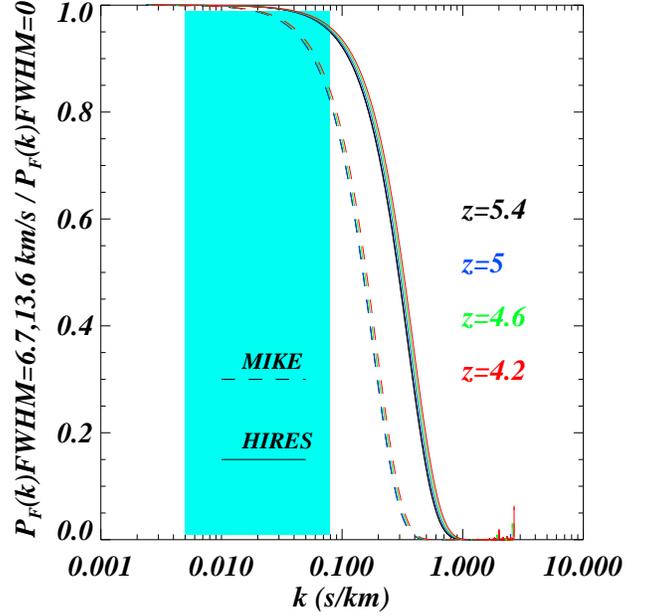}
\includegraphics[angle=0,width=9.cm,height=9.cm]{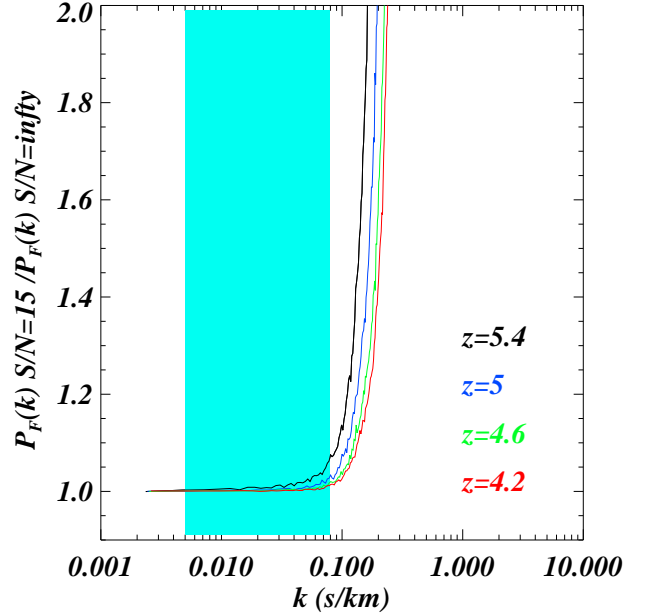}
\end{center}
\caption{\label{fig_h} {\it Upper panel:} The ratio of the 1D flux
  power spectrum for instrument resolutions corresponding to HIRES
  (6.7 km\,s$^{-1}$) and MIKE (13.6 km\,s$^{-1}$) spectrographs at 4 different
  redshifts ($z=4.2,4.6,5,5.4$) to the data with the native resolution
  of the simulation. {\it Lower panel:} The ratio of the 1D flux power
  spectrum for a S/N value per pixel equal to 15 to a model
  S/N=$\infty$.  The mock data have been corrected to incorporate both
  of these effects. The mean flux is the same in all the models shown,
  and the shaded area indicates the range of wavenumbers used in the
  present analysis.}
\end {figure} 

In Figure \ref{fig_h} we show the effect that instrumental resolution
and a given signal-to-noise (S/N) ratio has on the \lya flux power
spectrum in the four different redshift bins of our data sample.
These results have been obtained from our mock QSO spectra sample.
Note that both of these effects have been  incorporated  into the \lya forest spectra used in our
analysis. The S/N ratio results in an increase in the flux power of
less than 5\% over the range of wavenumbers considered in this work,
while instrumental resolution effects are particularly important for
the MIKE sample.  There is a 20\% correction at the smallest scales
for the MIKE data (for the HIRES data this value is 5\%).

\subsection{Systematic Effects on the Flux Power induced by Metals and UV background fluctuations}

\begin{figure}[h!]
\begin{center}
\includegraphics[angle=0,width=9.cm,height=9.cm]{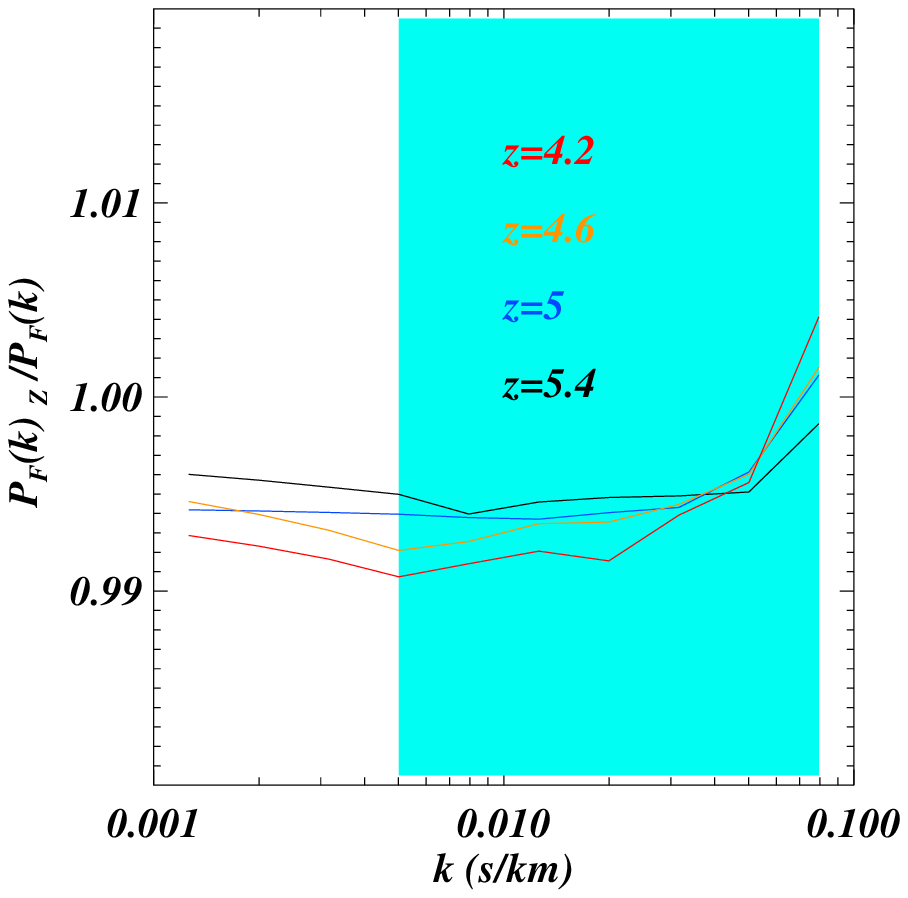}
\includegraphics[angle=0,width=9.cm,height=9.cm]{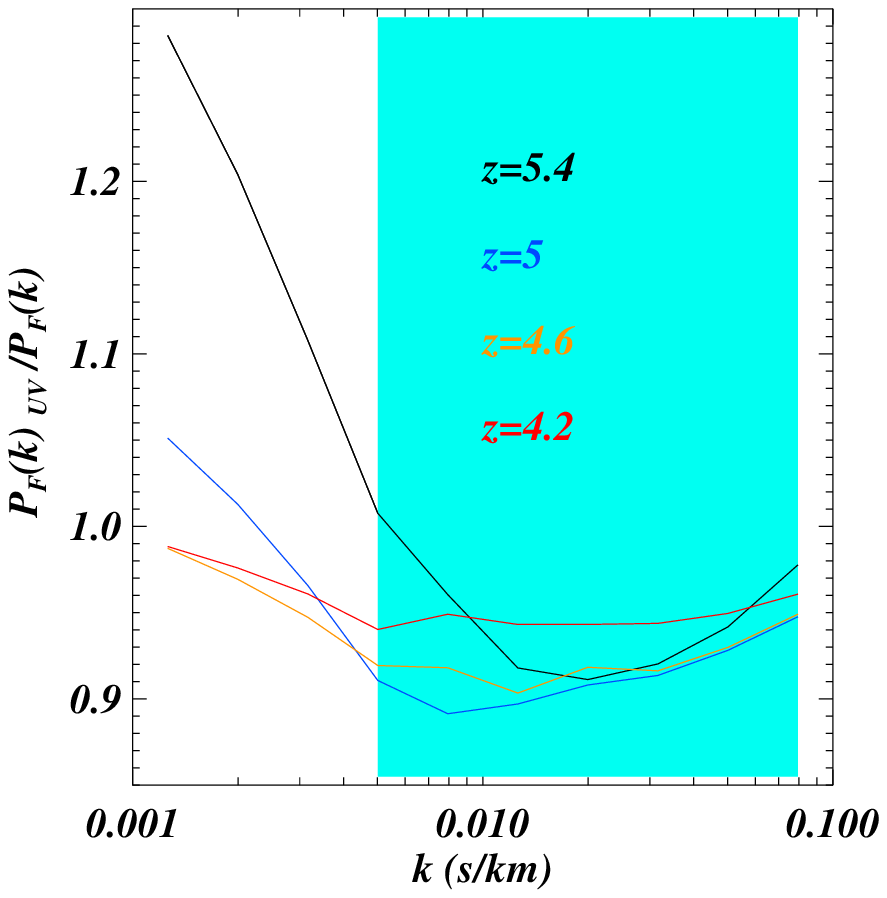}
\end{center}
\caption{\label{fig_hh}{\it Upper panel:} The ratio of the 1D flux
  power spectrum for a model including metal line contamination to the
  reference model at four different redshifts. {\it Lower panel:} The
  ratio of the 1D flux power spectrum for a model including the effect
  of spatial fluctuations of the  UV background at four different
  redshifts.  Note the different y-axis in the two figures.  The mean flux
  is the same in all the models shown, and the shaded area indicates
  the range of wavenumbers used in the present analysis.}
\end {figure} 

We also consider two important astrophysical nuissance effects in this work:
unwanted  contamination from metal lines in the \lya forest and
the effect of spatial fluctuations in the UV background intensity on
the observed \lya forest transmission.

In the redshift range we consider in this work, $z=4.2$--$5.4$, the
most common metal lines in the \lya forest will arise from absorbers
at lower redshifts.  We therefore consider the effect of absorption
from three prominent absorption line doublets; \CIV
($\lambda\lambda$1548, 1551), \SiIV ($\lambda\lambda$1394, 1403) and
\MgII ($\lambda\lambda$2796, 2804), arising over the redshift
intervals $z_{\rm CIV}=3.08$--$4.02$, $z_{\rm SiIV}=3.54$--$4.58$ and
$z_{\rm MgII}=1.26$--$1.78$, respectively.

We add these metal lines to \lya forest spectra drawn from our
reference $\Lambda$CDM hydrodynamical model using the following
prodcedure. We firstly integrate fits to the column density
distribution functions (CDDFs) presented by \cite{Scannapieco06} from
a set of 19 high resolution VLT/UVES quasar spectra at $z_{\rm
  qso}=2.1$--$3.3$ over the column density range $\log(N/\rm
cm^{-2})=12$-$15$, for all three species.  We then multiply the
results by the redshift path length of our \lya forest data set to
provide an estimate of the number of metal line absorbers in the \lya
forest.  Note that this approach will likely overestimate the number
of \CIV and \SiIV absorbers due to the somewhat lower redshift
coverage of the \cite{Scannapieco06} dataset relative to this work.
Our metal contamination estimates are therefore likely to be
conservative in this regard.  Next, we Monte Carlo sample column
densities and line widths for the appropriate number of lines from the
\cite{Scannapieco06} CDDF fits and the Doppler parameter distribution
given by Ref.~\cite{HuiRutledge99}, with a chosen value
$b_{\sigma}=10\rm\,km\,s^{-1}$ for all three species.  Finally, we
randomly insert these absorption features into the sight-lines in our
mock \lya forest data set.

In order to estimate the impact of spatial fluctuations in the UV
background on the \lya forest at $z=4.2$--$5.4$ we use a modified
version of the approach described in \cite{BoltonViel11}, which was
used to examine fluctuations in the \HeII ionising background at lower
redshift.  We refer the reader to Ref.~\cite{BoltonViel11} for further
details.  The key difference in this work is that we compute the
spatially varying \HI photo-ionisation rate along our simulated
sight-lines.  We achieve this by computing the specific intensity of
the ionising background between $1$--$4\rm\,Ry$ (replacing equation 3
in \cite{BoltonViel11}) by solving:

\begin{equation} 
\label{eq1} 
J({\bf r},\nu) = \frac{1}{4\pi}\sum_{i=1}^{N}\frac{L_{i}({\bf r}_{\rm i},\nu)}{4\pi|{\bf r}_{\rm i}-{\bf r}|^{2}}e^{-\frac{ |{\bf r}_{\rm i}-{\bf r}|}{\lambda_{\rm HI}}\left(\frac{\nu}{\nu_{\rm HI}}\right)^{-3(\beta-1)}}, 
\end{equation} 

\noindent
where $\nu_{\rm HI}$ is the frequency of the \HI ionisation edge,
$\lambda_{\rm HI}$ is the mean free path for ionising photons
presented by \cite{Songaila10} and $\beta=1.5$ is the power-law slope
of the \HI CDDF \cite{Kim02}.  The summation in eq.~\ref{eq1} is over all quasars
with luminosities, $L$, drawn from the \cite{Hopkins07} B-band quasar
luminosity function.  In order to be conservative, in this work we
adopt an extreme model which maximises the effect of the UV
fluctuations on the \lya forest by assuming all ionising photons in
the IGM at $z=4.2$--$5.4$ are produced by quasars with $M_B<-22$.
We therefore ignore the significant contribution to the UV background
from the more numerous, fainter star-forming galaxies at these
redshifts, which effectively smooth out the large-scale fluctuations
produced by the rarer quasars.  The spatially fluctuating
photo-ionisation rates are then obtained by integrating the specific
intensity with respect to frequency, weighted by the photo-ionisation
cross-section.

In Figure \ref{fig_hh} we show the effect that UV fluctuations and
metal contamination have on the flux power spectrum.  In the
wavenumber range considered here, the UV background fluctuations have
an effect at around the 10\% level at the largest scales, dropping to
5\% at smallest scales considered in this work. The effect is larger
at high redshift (z=5,5.4) than in the two other redshift bins.  Metal
contamination affects the flux power very little (at the $\pm 1\%$
level) and is at a level below the statistical error bars of our data.

\subsection{Mean flux level uncertainties}

\begin{figure}[h!]
\begin{center}
\includegraphics[angle=0,width=9.cm,height=9.cm]{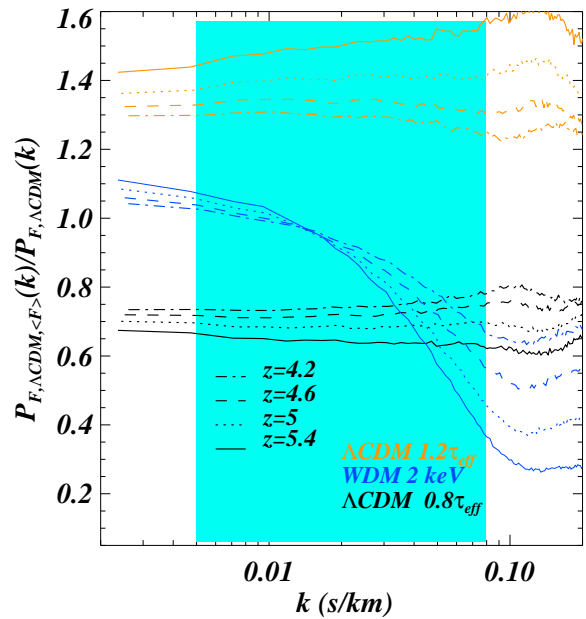}
\end{center}
\vspace{-0.5cm}
\caption{\label{fig_gg} The ratio of the 1D flux power spectrum for 2
  different $\Lambda$CDM models ($\tau_{\rm eff}=1.2\,\tau_{\rm
    eff,obs}$ in orange and $\tau_{\rm eff}=0.8\,\tau_{\rm eff,obs}$
  in black) at 4 different redshifts ($z=4.2,4.6,5,5.4$ represented by
  the dot-dashed, dashed, dotted and continuous curves, respectively)
  to the corresponding reference $\Lambda$CDM simulation. The WDM 2keV
  model is also shown in blue for comparison.  The shaded area
  indicates the range of wavenumbers used in the present analysis.}
\end {figure} 

The mean flux level $\langle F \rangle$, or alternatively the
effective optical depth $\tau_{\rm eff}=-\ln \langle F \rangle $, is a
key ingredient in our flux modeling procedure and a quantity that
needs to be marginalized over in the Monte Carlo Markov Chain
likelihood estmation.  In Figure \ref{fig_gg} we demonstrate the
effect that a different mean flux level has on the flux power
spectrum.  We choose two different mean flux levels with $\tau_{\rm
  eff}$ 20\% higher and lower than the reference value (which
corresponds to the observed mean flux).  A higher (lower) value for
$\tau_{\rm eff}$ will result in more (less) power relative to the
reference value. The trends are similar to those found for the
evolution of $\gamma$, i.e. rather flat in wavenumber space, with some
weak scale-dependence which is only present in the highest and lowest
redshift bins.  Our final results are not sensitive to the actual
choice of the effective optical depth values, since these are marginalized
over in the likelihood estimation.

\end{document}